\documentclass[useAMS,usenatbib, epsfig]{mn2e}   
\usepackage{amssymb}
\usepackage{amsmath}
\usepackage{amstext}
\usepackage{deluxetable}
\usepackage{epsfig}
\usepackage{thumbpdf}
\usepackage{epstopdf}
\usepackage{graphicx}
\usepackage[english]{babel}
\usepackage[utf8]{inputenc}
\usepackage{natbib}
\usepackage{xargs}                   
\usepackage[pdftex,dvipsnames]{xcolor} 
 \usepackage[colorinlistoftodos,prependcaption,textsize=tiny]{todonotes}
\usepackage[colorinlistoftodos,prependcaption,textsize=tiny]{todonotes}
\newcommandx{\unsure}[2][1=]{\todo[linecolor=red,backgroundcolor=red!25,bordercolor=red,#1]{#2}} 
\newcommandx{\change}[2][1=]{\todo[linecolor=blue,backgroundcolor=blue!25,bordercolor=blue,#1]{#2}} 
\newcommandx{\info}[2][1=]{\todo[linecolor=OliveGreen,backgroundcolor=OliveGreen!25,bordercolor=OliveGreen,#1]{#2}} 
\newcommandx{\improvement}[2][1=]{\todo[linecolor=Plum,backgroundcolor=Plum!25,bordercolor=Plum,#1]{#2}}
\newcommandx{\thiswillnotshow}[2][1=]{\todo[disable,#1]{#2}}
\usepackage{lmacs}
\usepackage{savesym}

\usepackage[citecolor=blue]{hyperref}
\hypersetup{colorlinks=true}
 \usepackage{pdf14}
 
\DeclareRobustCommand{\ion}[2]{%
\relax\ifmmode
\ifx\testbx\f@series
{\mathbf{#1\,\mathsc{#2}}}\else
{\mathrm{#1\,\mathsc{#2}}}\fi
\else\textup{#1\,{\mdseries\textsc{#2}}}%
\fi}

\title[Spectroscopic characterisation of SN 2014J] {SN 2014J at M82: I. 
A middle-class type Ia supernova by all spectroscopic metrics}

\author[Galbany et al.]{L. Galbany$^{1,2}$\thanks{E-mail: lgalbany@das.uchile.cl},
M.~E. Moreno-Raya$^{3}$,
P. Ruiz-Lapuente$^{4,5}$,
J.~I. Gonz\'alez Hern\'andez$^{6,7}$,
\newauthor
J. M\'endez$^{8}$,
P. Vallely$^{9}$,
E. Baron$^{9}$,
I. Dom\'inguez$^{10}$,  
M. Hamuy$^{2,1}$
\newauthor
A.~R. L\'opez-S\'anchez$^{11,12}$,
M. Moll\'a$^{3}$,
S. Catal\'an$^{13}$,
E. A. Cooke$^{14}$,
C. Fari\~na$^{8}$,
\newauthor
R. G\'enova-Santos$^{6,7}$,
R. Karjalainen$^{8}$,
H. Lietzen$^{6,7}$,
J. McCormac$^{13}$,
F.~C. Riddick$^{8}$,
\newauthor
J. A. Rubi\~no-Mart{\'\i}n$^{6,7}$,
I. Skillen$^{8}$,
V. Tudor$^{8}$,
O. Vaduvescu$^{8}$\\
$^1$Millennium Institute of Astrophysics, Universidad de Chile, Chile\\ 
$^2$Departamento de Astronom\'ia, Universidad de Chile, Santiago, Chile\\
$^3$Departamento de Investigaci\'on B\'asica, CIEMAT, Avda. Complutense 40, 28040, Madrid, Spain\\
$^4$Instituto de F\'isica Fundamental, Consejo Superior de Investigaciones Cient\'ificas, c/. Serrano 121, E-28006, Madrid, Spain \\
$^5$Institut de Ci\`encies del Cosmos (UB-IEEC), c/. Mart\'i i Franqu\'es 1, E-08028, Barcelona, Spain \\
$^6$Instituto de Astrof\'isica de Canarias, 38200 La Laguna, Tenerife, Spain \\
$^7$Departamento de Astrof\'isica, Universidad de La Laguna (ULL), 38206 La Laguna, Tenerife, Spain\\
$^8$Isaac Newton Group of Telescopes, Apto. 321, E-38700 Santa Cruz de la Palma, Canary Islands, Spain\\
$^9$Homer L. Dodge Department of Physics and Astronomy, University of Oklahoma, 440 W. Brooks, Rm 100, Norman, OK 73019-2061, USA \\
$^{10}$Universidad de Granada, E-18071, Granada, Spain \\
$^{11}$Australian Astronomical Observatory, P.O. Box 915, North Ryde, NSW 1670, Australia\\
$^{12}$Department of Physics and Astronomy, Macquarie University, NSW 2109, Australia\\
$^{13}$Department of Physics, University of Warwick, Gibbet Hill Road, Coventry, CV4 7AL, UK\\
$^{14}$School of Physics and Astronomy, University of Nottingham, University Park, Nottingham, NG7 2RD, UK.\\
$^{15}$Centro de Astrobiolog\'ia (CSIC-INTA), Ctra. Ajalvir km 4, E-28850, Torrej\'on de Ardoz, Madrid, Spain
}

\date{Received date: October 22, 2015; accepted date: January 4th, 2016}

\begin{document}
\maketitle
\begin{abstract}
We present the intensive spectroscopic follow up of the type Ia supernova (SN Ia) 2014J in the starburst galaxy M82.
Twenty-seven optical spectra have been acquired from January 22$^{\rm nd}$ to September 1$^{\rm st}$ 2014 with the Isaac Newton (INT) and William Herschel (WHT) Telescopes. 
After correcting the observations for the recession velocity of M82 and for Milky Way and host galaxy extinction, we measured expansion velocities from spectral line blueshifts and pseudo-equivalent width of the strongest features in the spectra, which gives an idea on how elements are distributed within the ejecta. 
We position SN 2014J in the Benetti (2005), Branch et al. (2006) and Wang et al. (2009) diagrams.
These diagrams are based on properties of the Si II features and provide dynamical and chemical information about the SN ejecta. 
The nearby SN 2011fe, which showed little evidence for reddening in its host galaxy, is shown as a reference for comparisons. 
SN 2014J is a border-line object between the Core-normal ({\sc CN}) and Broad-line ({\sc BL}) groups, which corresponds to an intermediate position between Low Velocity Gradient ({\sc LVG}) and High Velocity Gradient ({\sc HVG}) objects. 
SN 2014J follows the $R({\rm Si~II})$--$\Delta$m$_{15}$ correlation, which confirms its classification as a relatively normal SN Ia. 
Our description of the SN Ia in terms of the evolution of the pseudo-equivalent width of various ions as well as the position in the various diagrams put this specific SN Ia into the overall sample of SN Ia.
\end{abstract}

\begin{keywords}
techniques: spectroscopic; (stars:) supernovae: general; (stars:) supernovae: individual: 2014J; methods: data analysis  
\end{keywords}
 

\section{Introduction} \label{Section1}

Type Ia supernovae (SN Ia) are close binary systems where one of the stars, a carbon-oxygen white dwarf (C+O WD), undergoes a thermonuclear runaway following the start of explosive burning at its center  \citep{1960ApJ...132..565H}. The physics of the explosion is determined by both components of the system. While the ejecta from the explosion (the supernova remnants, SNRs) have been studied in great detail, little is known yet about the companion star. It could be another WD (the DD channel, \citealt{1984ApJS...54..335I}) or a star still fueled by thermonuclear burning (the SD channel, \citealt{1973ApJ...186.1007W}; it includes the case where the explosion is due to the merging of a WD with the core of an AGB star, proposed by \citealt{2013MNRAS.431.1541S, 2003ApJ...594L..93L}). See for recent reviews \cite{2013FrPhy...8..116H}, \cite{2014ARA&A..52..107M}, and \cite{2014NewAR..62...15R}.

\begin{figure*}
\centering
\includegraphics[width=\linewidth]{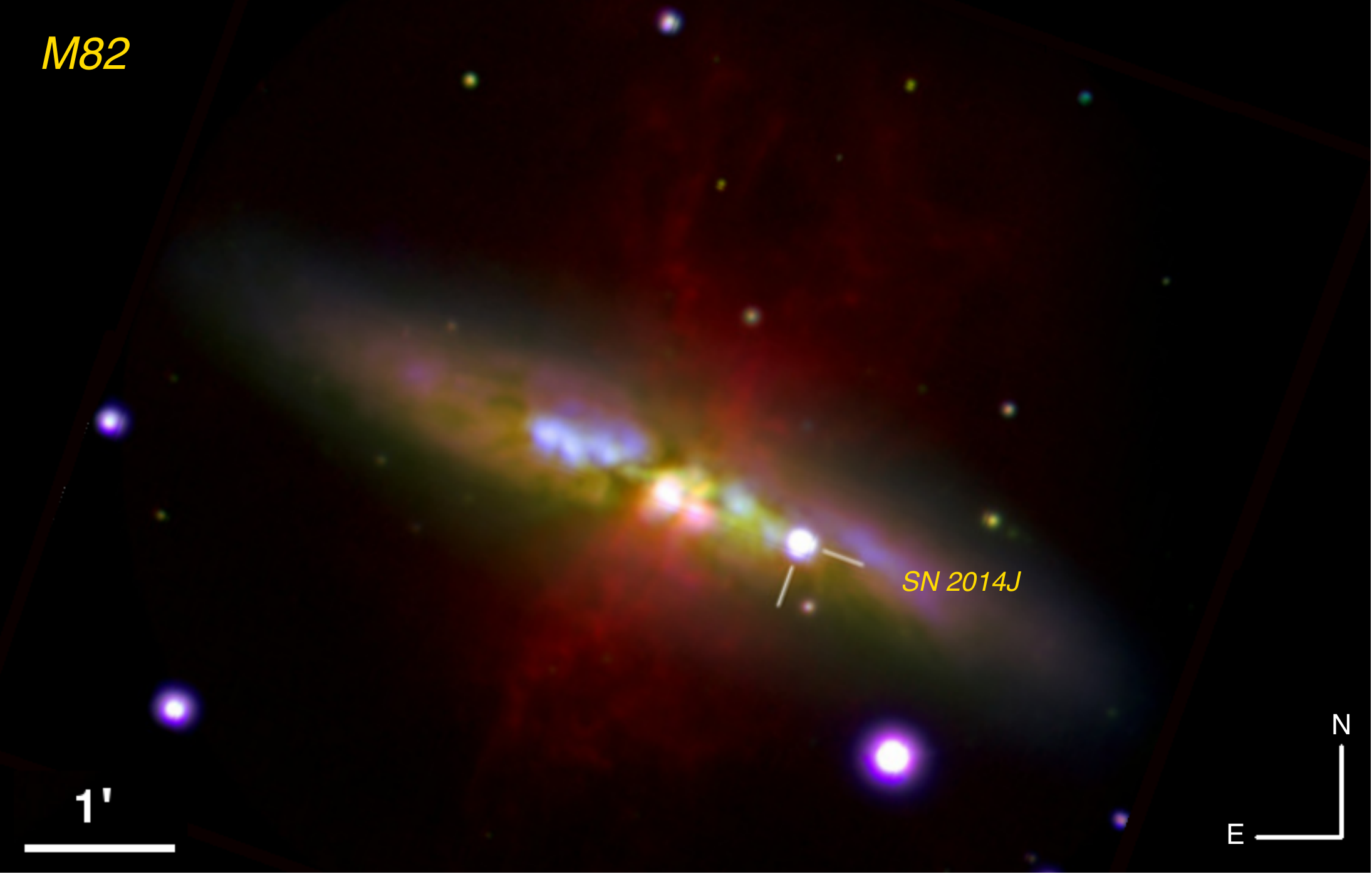}     
\caption{Type Ia SN 2014J in starburst galaxy M82. Image obtained using ACAM at 4.2m William Herschel Telescope combining the following observations: 2$\times$300 s in $u$ (dark blue), 3$\times$100 s in $g$ (cyan),  3$\times$200 s in $i$ (green),  3$\times$100 s in $r$ (magenta), and 4$\times$300 s in H$\alpha$ band filter (red).}
\label{M82}
\end{figure*}

A few SN Ia per century occur in a typical galaxy, hence the probability of finding a nearby object is very low \citep{2011MNRAS.412.1473L}. Therefore, the discovery of the nearby SN 2014J has become an excellent opportunity to improve the understanding on the supernova physics, by studying with in detail and with relatively high spatial resolution the exploding star and its environment.
In addition, precise characterization of their observed heterogeneity would be beneficial for reducing the systematic uncertainties of SN Ia distance method for cosmology.

SN 2014J exploded in the edge-on galaxy M82 (at $\alpha$ = 09$^h$55$^m$42.11$^s$, $\delta$ = +69$^\circ$40'25.87'', see Figure \ref{M82}) which is located only $\sim$3.5 Mpc away \citep{2009ApJS..183...67D}.
It is definitely the nearest SN Ia ever observed using modern instrumentation such as CCDs (since SN 1987A was a peculiar type II SN), surpassing SN 2011fe another SN Ia found in M101 ($d$ = 6.4 $\pm$ 0.7 Mpc). It rivals SN 1972E for being the closest SN Ia discovered in the last 4 decades, considering the large uncertainty in NGC 5253 distance estimation (SN 1972's host galaxy).
SN 2014J was discovered serendipitously by \cite{2014CBET.3792....1F} during a lesson on January 21.81 UT, while using the 0.35m telescope at University of London Observatory. 
After Fossey reported the discovery some pre-discovery observations were advertised \citep{2014ATel.5794....1M, 2014ATel.5795....1D}.

SN 2014J was first classified as a SN Ia by \cite{2014ATel.5786....1C} from a spectrum obtained with the Dual Imaging Spectrograph on the ARC 3.5m telescope. 
As the nearest modern SN~Ia, SN~2014J has been extremely well followed-up by many groups, with different instruments, and in several wavelength ranges: 
$\gamma$-rays \citep{2014Natur.512..406C, 2014Sci...345.1162D, 2015AN....336..464D}, 
X-rays \citep{2014ApJ...790...52M}, 
UV \citep{2015ApJ...805...74B},
optical \citep{2014ATel.5816....1K, 2014ATel.5827....1M, 2014ATel.5829....1B, 2014ATel.5843....1H, 2015ApJ...799..197R, 2015ApJ...799..105S, 2015JAVSO..43...43P}, 
near-IR \citep{2014ATel.5840....1R, 2014ApJ...784L..12G, 2014ATel.5876....1S, 2015ApJ...798...39M, brian_14JIR14, 2015ApJ...804...66V} 
mid-IR \citep{2015ApJ...798...93T}, 
radio \citep{2015arXiv151007662C,2014ATel.5804....1C, 2014ATel.5812....1C, 2014ApJ...792...38P, 2014ATel.6197....1S}, 
and polarimetry \citep{2014ApJ...795L...4K,2015A&A...577A..53P}.

A few weeks after the discovery the first analyses of this object were presented.
The epoch of explosion was constrained to be around January 14.75 $\pm$ 0.30 UT (56671.75 MJD, \citealt{2014ApJ...783L..24Z, 2015ApJ...799..106G}), and a maximum brightness in the $B$-band of $M_B=-19.19~\pm~0.10$ mag, was reached on February $1.74 \pm 0.13$ UT (56689.74 MJD, \citealt{2015ApJ...798...39M}).
Using optical photometry \cite{2014CoSka..44...67T} estimated a decline rate parameter of $\Delta $m$_{15}$ = 1.01 $\pm$ 0.05 mag, while \cite{2014MNRAS.445.4427A} reported 1.08 $\pm$ 0.03 mag, \cite{2015ApJ...798...39M} 1.12 $\pm$ 0.02 mag, and \cite{2014ApJ...795L...4K} 1.02 $\pm$ 0.05 mag, pointing out that SN 2014J light curve has a typical stretch factor of a normal SN Ia.
Due to the inclination with respect the line-of-sight of M82, SN 2014J has shown to be highly affected by dust extinction.
\cite{2014ApJ...788L..21A} and \cite{2014MNRAS.443.2887F} presented extensive independent studies on extinction modeling, both arriving at the conclusion that an $R_V$ (1.3 - 2.0) lower than the standard Galactic value ($\sim$ 3.1) is needed to explain the observations, which is in agreement with the results from other work \citep{2014ApJ...784L..12G, 2015ApJ...805...74B, 2015ApJ...807L..26G}, and confirmed with polarimetry by \cite{2014ApJ...795L...4K} favoring dust grains with smaller radii and different nature than the typical Galactic dust.
Radio and X-ray observations reported no detection of the SN in these wavelengths which was interpreted as a support for DD scenario for SN 2014J, and was also used to estimate the progenitor mass-loss rate \citep{2014ApJ...792...38P, 2014ApJ...790...52M, 2014MNRAS.442.3400N}.
This estimate is in agreement with the upper limits estimated by \cite{2014ApJ...790....3K} using HST data from near-UV to near-NIR  and by \cite{2015A&A...577A..39L} using late optical spectra (but see \citealt{2015ApJ...801..136G} for evidence of multiple components of CSM which would favor the SD scenario, and \cite{2015PASJ...67...54K} who do not rule it out). 
Interestingly, \cite{2015ApJ...804L..37C} reported the discovery of light-echoes at $\sim$80 pc from the SN location revealing the CSM structure. 
Although all previous works have reported measurements of several properties that are in agreement with those of typical SN Ia, a proper spectral characterization and comparison to the whole SN Ia population has not been yet performed.

\begin{table*}\scriptsize
\caption{Instrument specifications for the observations presented in this work.}
 \label{tab:obs}
\centering        
\begin{tabular}{lcccccccccc}
\hline\hline   
Date  &Epoch &Telescope &Instrument &Grism/ &Camera &Exp. time &Spatial disp.   &Spectral disp.      &$\lambda$ range &Slit width \\
 (UT) & (days) &                &                   &  grating        &  detector    & (s)           &("~pix$^{-1}$)&(\AA~pix$^{-1}$)& (\AA)                    & (") \\
\hline                                                       
JAN 23.18 & -9.6   & INT & IDS  & R1200R & EEV10    & 4x300     & 0.40 & 0.47 & 5800-7400 & 1.0 \\
JAN 24.23 & -8.5   & WHT & ACAM & V400   & AUXCAM   & 3x60      & 0.25 & 3.30 & 3500-9400 & 1.0 \\
JAN 25.30 & -7.4   & WHT & ACAM & V400   & AUXCAM   & 2x200     & 0.25 & 3.30 & 3500-9400 & 1.0 \\
          &        & WHT & ISIS & R600B  & EEV12    & 2x900     & 0.20 & 0.45 & 3526-5350 & 1.0 \\
          &        & WHT & ISIS & R600R  & REDPLUS  & 2x900     & 0.22 & 0.49 & 4872-6926 & 1.0 \\
          &        & WHT & ISIS & R600R  & REDPLUS  & 2x900     & 0.22 & 0.49 & 5757-7811 & 1.0 \\
JAN 26.25 & -6.5   & WHT & ACAM & V400   & AUXCAM   & 3x200     & 0.25 & 3.30 & 3500-9400 & 1.0 \\
JAN 27.30 & -5.4   & WHT & ACAM & V400   & AUXCAM   & 2x200     & 0.25 & 3.30 & 3500-9400 & 1.0 \\
          &        & WHT & ISIS & R1200B & EEV12    & 2x1200    & 0.20 & 0.45 & 3480-4420 & 1.0 \\
          &        & WHT & ISIS & R1200R & REDPLUS  & 3x600     & 0.22 & 0.49 & 5361-6416 & 1.0 \\
JAN 28.11 & -4.6   & WHT & ACAM & V400   & AUXCAM   & 2x15	& 0.25 & 3.30 & 3500-9400 & 2.0 \\
FEB 04.05 & +2.3   & INT & IDS  & R1200Y & REDPLUS2 & 3x120	& 0.44 & 0.53 & 4900-6400 & 1.3 \\
          &        & INT & IDS  & R1200Y & REDPLUS2 & 3x90	& 0.44 & 0.53 & 5860-7400 & 1.3 \\
FEB 05.06 & +3.3   & WHT & ISIS & R600B  & EEV12    & 2x60	& 0.20 & 0.45 & 3700-5050 & 1.0 \\
	  &        & WHT & ISIS & R600R  & REDPLUS  & 2x60	& 0.22 & 0.49 & 5600-7150 & 1.0 \\
FEB 05.93 & +4.2   & WHT & ISIS & R600B  & EEV12    & 2x120	& 0.20 & 0.45 & 3650-5050 & 1.0 \\
	  &        & WHT & ISIS & R600R  & REDPLUS  & 2x20	& 0.22 & 0.49 & 5650-7200 & 1.0 \\
FEB 12.27 & +10.5  & INT & IDS  & R300V  & REDPLUS2 & 7x60	& 0.44 & 2.06 & 4700-9900 & 1.2 \\
FEB 14.10 & +12.4  & INT & IDS  & R400V  & REDPLUS2 & 6x120	& 0.44 & 1.55 & 4400-7400 & 8.0 \\
FEB 19.18 & +17.4  & WHT & ACAM & V400   & AUXCAM   & 3x30	& 0.25 & 3.30 & 4900-9300 & 1.0 \\
FEB 20.07 & +18.3  & WHT & ACAM & V400   & AUXCAM   & 3x30	& 0.25 & 3.30 & 4900-9300 & 1.0 \\
FEB 21.20 & +19.5  & WHT & ACAM & V400   & AUXCAM   & 3x30	& 0.25 & 3.30 & 4900-9300 & 1.0 \\
FEB 24.19 & +22.5  & WHT & ACAM & V400   & AUXCAM   & 7x30	& 0.25 & 3.30 & 4900-9300 & 1.0 \\
FEB 26.27 & +24.5  & WHT & ACAM & V400   & AUXCAM   & 1x30	& 0.25 & 3.30 & 4900-9300 & 1.0 \\
MAR 07.18 & +33.4  & WHT & ACAM & V400   & AUXCAM   & 2x30	& 0.25 & 3.30 & 4900-9300 & 1.0 \\
MAR 08.16 & +34.4  & WHT & ACAM & V400   & AUXCAM   & 1x30	& 0.25 & 3.30 & 4900-9300 & 1.0 \\
MAR 12.86 & +39.1  & WHT & ACAM & V400   & AUXCAM   & 1x30	& 0.25 & 3.30 & 4100-9200 & 1.0 \\
MAR 18.14 & +44.4  & WHT & ACAM & V400   & AUXCAM   & 1x30	& 0.25 & 3.30 & 4100-9200 & 1.0 \\
APR 08.91 & +66.2  & WHT & ACAM & V400   & AUXCAM   & 1x60	& 0.25 & 3.30 & 4900-9300 & 1.0 \\
MAY 08.91 & +96.2  & WHT & ACAM & V400   & AUXCAM   & 1x30	& 0.25 & 3.30 & 4900-9300 & 1.0 \\
MAY 09.90 & +97.2  & WHT & ISIS & R300B  & EEV12    & 3x60	& 0.20 & 0.86 & 3700-5350 & 1.0 \\
	  &        & WHT & ISIS & R316R  & REDPLUS  & 3x60	& 0.22 & 0.93 & 5350-8000 & 1.0 \\
MAY 11.89 & +99.1  & WHT & ISIS & R600B  & EEV12    & 3x30	& 0.20 & 0.45 & 3700-5350 & 1.0 \\
	  &        & WHT & ISIS & R600R  & REDPLUS  & 3x60	& 0.22 & 0.49 & 6200-7700 & 1.0 \\
JUN 18.93 & +137.2 & WHT & ISIS & R158B  & EEV12    & 1x300	& 0.20 & 1.62 & 3650-5050 & 1.0 \\
          &        & WHT & ISIS & R158R  & REDPLUS  & 1x300	& 0.22 & 1.81 & 5500-9050 & 1.0 \\
JUL 08.89 & +157.2 & WHT & ACAM & V400   & AUXCAM   & 1x120	& 0.25 & 3.30 & 4900-9300 & 1.0 \\
SEP 02.23 & +212.5 & WHT & ACAM & V400   & AUXCAM   & 1x30      & 0.25 & 3.30 &	4900-9300 & 1.0 \\
\hline
\end{tabular}
\end{table*}

In the framework of a dedicated program guaranteed by the Isaac Newton Group (ING) we obtained a long time baseline set of optical spectroscopy, with wavelengths from roughly 3500 {\AA} to 9500 {\AA}, with the 2.5m \textit{Isaac Newton} ({\sc INT}) and 4.2m \textit{William Herschel} ({\sc WHT}) telescopes, both located at the El Roque de los Muchachos Observatory, La Palma. The spectral epochs range from almost two weeks pre-maximum (January 22$^{\rm nd}$ 2014) to nearly seven months post-maximum (September 1$^{\rm st}$ 2014). A total of 27 different epochs were acquired with different instruments, technical configurations, and spectral resolutions.
In addition, several broad-band images were also taken on different epochs. In Figure \ref{M82} we show a false color image of the SN 2014J and its host galaxy M82, composed from our observations at the WHT using $ugri$ and H$\alpha$ filter images.

In this work we present these observations (in section \S\ref{Section2}), and describe the characterization of SN 2014J in the spectral SN Ia diagrams of \cite{2005ApJ...623.1011B} [hereafter BE05],  \cite{2006PASP..118..560B} [hereafter BR06], and \cite{2009ApJ...699L.139W}  [hereafter WA09], together with the evolution of the velocity and pseudo-equivalent width of several spectral features (See \citealt{2014Ap&SS.351....1P} for an extended review on SN Ia spectroscopy). We also use SN 2011fe as a reference for comparisons. Those diagrams provide information on the dynamics and chemical distribution of SN Ia events, and have been of great utility to characterize large samples of SN Ia such as the CfA sample \citep{2012AJ....143..126B} or the Carnegie Supernovae Project (CSP, \citealt{2013ApJ...773...53F}). We add this very well studied SN Ia into the overall samples for a better understanding of this event (sections \S\ref{Section3} and \S\ref{Section4}). Finally, in section \S\ref{Section5}, we give a summary and conclusions.
A comparison with synthetic spectra using SYNOW \citep{2007PASP..119..709B} is presented in Paper II \citep{2015arXiv151202608V}.


\section{Observations and reduction} \label{Section2}

We obtained long-slit spectroscopy of SN 2014J using the 2.5m {\sc INT} and 4.2m {\sc WHT} telescopes, both located at the El Roque de los Muchachos Observatory on La Palma. Spectra were obtained on 27 nights from January 22$^{\rm nd}$ to September 1$^{\rm st}$ 2014 in different ways: observation time was allocated to the ING service proposal {\it SW2014a08: Spectroscopic follow-up of SN 2014J}, (PI: P. Ruiz-Lapuente), ING discretionary time on the INT, and time offered by several programs both at WHT and INT. 
Since both the observers and the configuration of the instruments changed every night, we describe separately the different setups for each spectrograph. Details for individual spectra are given in Table \ref{tab:obs}.

\subsection{Observations using 2.5m INT}

Four spectra (on nights Jan 22$^{\rm nd}$, Feb 03$^{\rm rd}$, 11$^{\rm th}$ and 13$^{\rm th}$) were obtained using the Intermediate Dispersion Spectrograph ({\sc IDS}) mounted at the Cassegrain focus of the {\sc INT}. Two detectors were used depending on the set up in different nights.
On Jan 22$^{\rm nd}$ the EEV10 detector was used, which has a pixel size of 13.5 $\mu$m and a spatial dispersion of 0.40$^{\prime\prime}$~pix$^{-1}$. 
The observation was performed with the R1200R grating, with a spectral dispersion of 0.47 \AA~ pix$^{-1}$, and a 1$^{\prime\prime}$-width slit. 
On the other three nights the REDPLUS2 detector, with a pixel size of 15 $\mu$m and a spatial dispersion of 0.44$^{\prime\prime}$~pix$^{-1}$, was used.
Observations were performed with three different gratings: R1200Y (spectral dispersion of 0.53 \AA~ pix$^{-1}$) and the 1.3$^{\prime\prime}$ slit; R300V (spectral dispersion of 2.06 \AA~ pix$^{-1}$) with the 1.2$^{\prime\prime}$ slit; and R400V (spectral dispersion of 1.55 \AA~ pix$^{-1}$) with the 8$^{\prime\prime}$ slit.

\subsection{Observations using 4.2m WHT}

The remaining spectra were taken at the 4.2m {\sc WHT} using either the Intermediate dispersion Spectrograph and Imaging System ({\sc ISIS}) or the Auxiliary-port CAMera ({\sc ACAM}).

ISIS is mounted at the Cassegrain focus and consists of a dual-beam spectrograph with two independent arms which provide simultaneous observations in blue and red bands of the spectrum. The blue arm incorporates an EEV12 detector, with a pixel size of 13.5 $\mu$m, and spatial dispersion of  0.20$^{\prime\prime}$~pix$^{-1}$. 
The gratings used were: R600B (spectral dispersion of 0.45 \AA~pix$^{-1}$), R300B (spectral dispersion of 0.86 \AA~ pix$^{-1}$) and R158B (spectral dispersion of 1.62 \AA~pix$^{-1}$). Slit width was 1$^{\prime\prime}$ in all cases. 
The red arm has a REDPLUS detector, with a pixel size of 15.0 $\mu$m and a spatial dispersion of  0.22$^{\prime\prime}$~pix$^{-1}$. In this arm the configurations used have the following specifications: R600R (spectral dispersion 0.49 \AA~ pix$^{-1}$); R316R (spectral dispersion 0.93 \AA~ pix$^{-1}$); and R158R (spectral dispersion 1.81 \AA~ pix$^{-1}$). Slit width was also 1$^{\prime\prime}$ in all cases. 

ACAM is mounted permanently at a folded-Cassegrain focus, and it provides fixed-format low-resolution spectroscopy. All spectra were taken with the 400-line grism V400, which covers a spectral range from $\sim$ 3500 to $\sim$ 9400 \AA~ ($\sim$ 3.3 \AA~pix$^{-1}$), and it has a spatial dispersion of 0.25$^{\prime\prime}$ pix$^{-1}$. Slit width was 1$^{\prime\prime}$ for all cases except one (Jan 28$^{\rm th}$) when the 2.0$^{\prime\prime}$ wide slit was used. 

\begin{figure*}
\centering
\includegraphics[width=\linewidth]{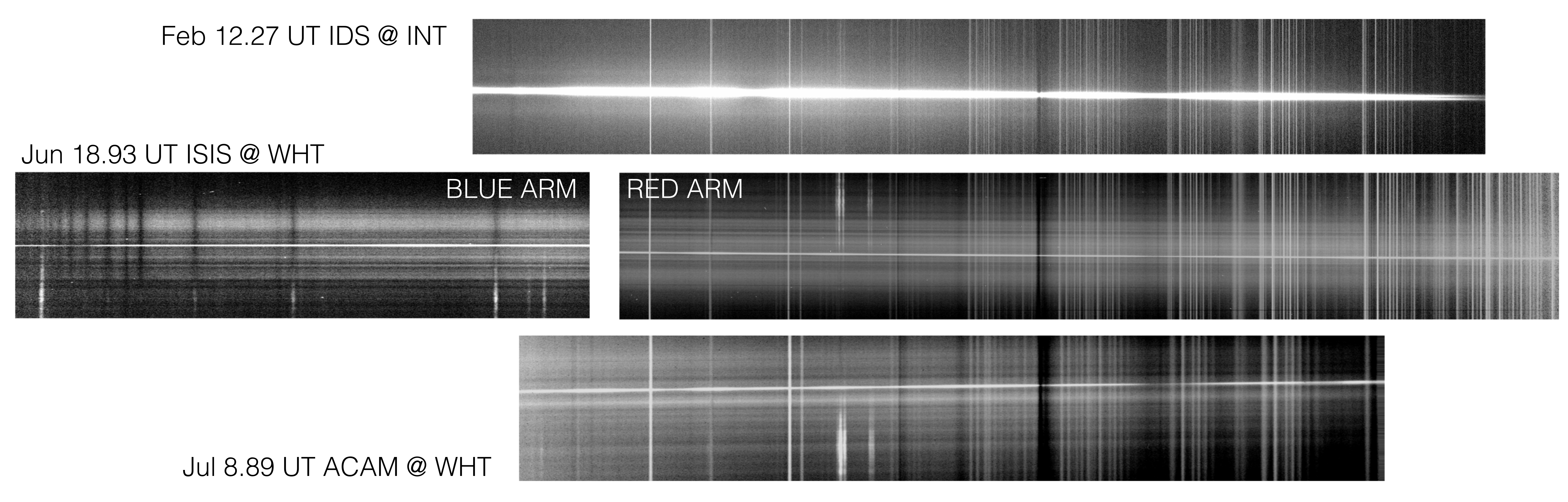}   
\caption{2D spectra of SN 2014J taken with different instruments at both telescopes. Top row: spectrum taken at INT using IDS covering the spectral range 4700-9900 \AA~with R300V. Mid row: blue and red arm spectra taken with ISIS at WHT, taken with the R600B and R600R gratings, respectively.. The wavelength coverage range from 3650-5050 \AA~in the blue arm and from 5500-9050 \AA~in the red arm. Bottom row: ACAM spectrum covering the optical range 4900-9300 \AA. Several emission lines from the host galaxy M82 can be seen in the mid and bottom spectra. }
\label{ds9spectra}
\end{figure*}

\subsection{Reduction} \label{sec:red}

All spectra have been reduced using standard {\sc IRAF}\footnote[1]{Image Reduction Analysis Facility, distributed by the National Optical Astronomy Observatories (NOAO), which is operated by AURA Inc., under cooperative agreement with NSF} routines, including debiasing and flat-fielding. The spectra have been calibrated in wavelength taking observations of internal arc lamps installed in the Acquisition and Guiding (A\&G) Box using the same instrumental configuration as for the science observations. The following standard stars were observed for the flux calibration: HD93521, BD08 2015, Grw+70 5824, G191-B2B, Feige 34, BD+75 325, PG0934+554 belonging to \cite{1988ApJ...328..315M}, \cite{1990AJ.....99.1621O} and \cite{1977ApJ...218..767S} catalogues; and HD109995, all included in the standard database of {\sc IRAF}. 
The routines used for these purposes were: {\sc identify}, {\sc reidentify}, {\sc fitcoords} and {\sc transform} (for calibration in wavelength); and {\sc standard}, {\sc sensfunc} and {\sc calibrate} (for flux calibration).
In Figure \ref{ds9spectra} we show examples of three 2-dimensional spectra from different instruments at both telescopes, where the wavelength coverage, strong emission lines, and sky contamination can be seen.
Figure~\ref{27spectra} shows a composite of the fully-reduced observed spectra, where all have been shifted to the rest frame by the recession velocity of M82 ($v$ = 203 km s$^{-1}$). 
The complete set of spectroscopy is available electronically\footnote{\href{https://github.com/lgalbany/SN2014J}{https://github.com/lgalbany/SN2014J}} or can be downloaded from the Weizmann interactive supernova data repository (WISeREP\footnote{\href{http://wiserep.weizmann.ac.il}{http://wiserep.weizmann.ac.il}}; \citealt{2012PASP..124..668Y}).

 
\section{Spectral characterization}  \label{Section3}

\begin{figure*} 
\centering
\includegraphics[trim=2.0cm 0.4cm 0.3cm 0.9cm, clip=true,width=0.94\linewidth]{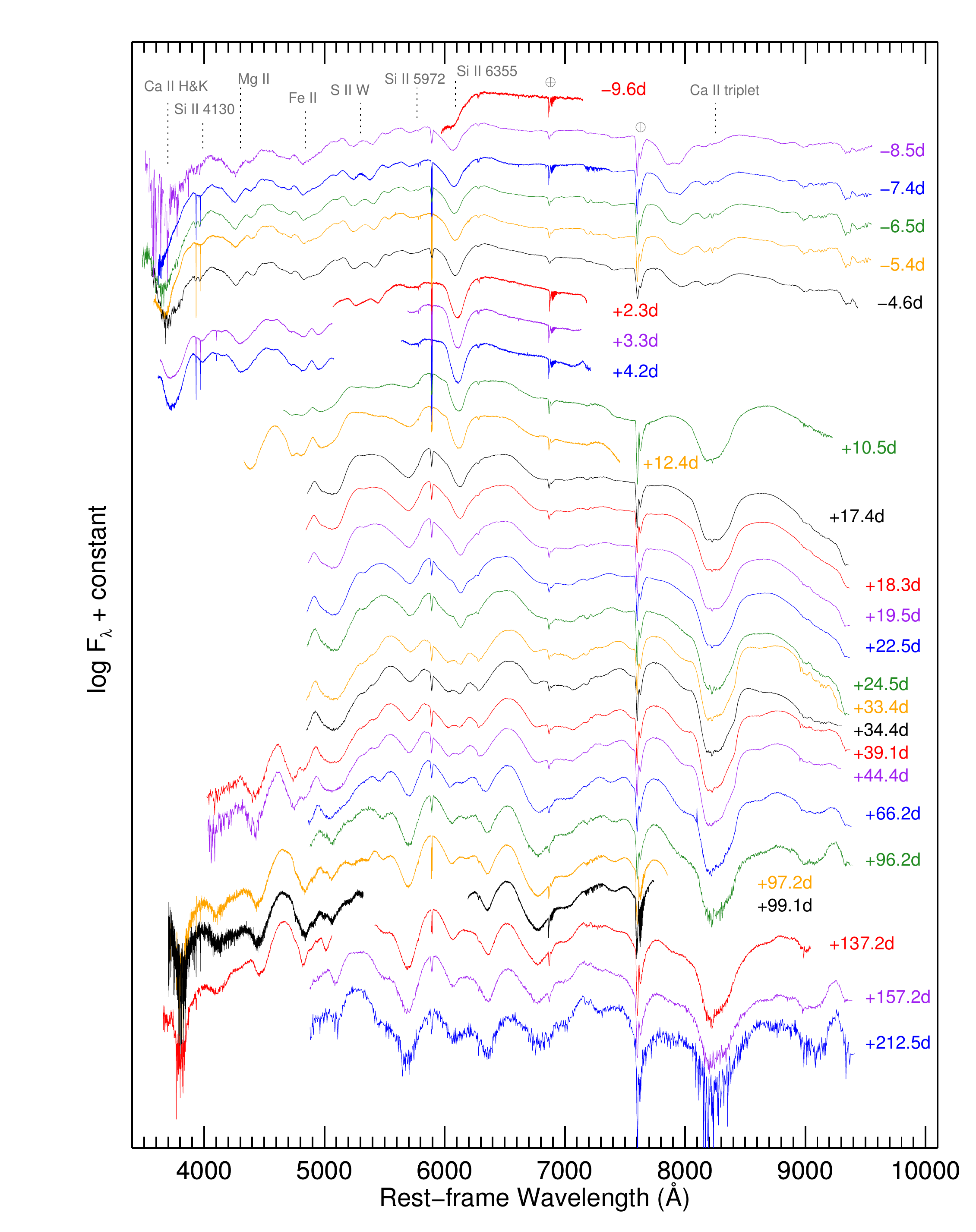}   
\caption{Spectral evolution of SN 2014J. Observed unredshifted pre-maximum, maximum-light, and post-maximum spectra taken at ORM (La Palma). The maximum light in the $B$-band has been determined to be at February $1.74 \pm 0.13$ UT (56689.74 MJD). For epochs -7.4d and -5.4d from maximum, 3 and 2 respectively high-resolution spectra have also been observed, and here we show a composite of these and the low-res spectra.} 
\label{27spectra}
\end{figure*}

The evolution of several spectral features is clearly seen in Figure~\ref{27spectra}, including the typical features found in SN Ia spectra: Ca II H\&K,  Si II $\lambda$4130, Mg II, S II W, Si II $\lambda$5972, Si II $\lambda$6355, and Ca II triplet.
There is also evidence of high-velocity features (e.g. HV Ca II at $\sim$7900 \AA~in the pre-maximum spectra).
Telluric lines, marked with Earth symbols, and ISM absorptions from the host galaxy, such as NaD $\lambda$5900, have not been removed in the Figure but in the measurements when necessary.
From bluer wavelengths we see, as pointed out by previous works, that the SN is heavily reddened by dust in the host galaxy. 
Although the spectra shown in the Figure has not been corrected for Milky Way (MW) and host galaxy extinction, the spectral parameters described below have been measured after applying these corrections using a \citealt{1999PASP..111...63F} law, for both the MW using the dust maps of \cite{2011ApJ...737..103S} assuming an $R_V=3.1$, and for the host galaxy with the reported values of $E(B-V)=1.2$ and $R_V=1.4$ from \cite{2014ApJ...784L..12G}.

\subsection{Measurement of the spectral parameters}

\begin{figure}
\centering
\includegraphics[trim=0.9cm 0.3cm 0.3cm 0.6cm, clip=true,width=\linewidth]{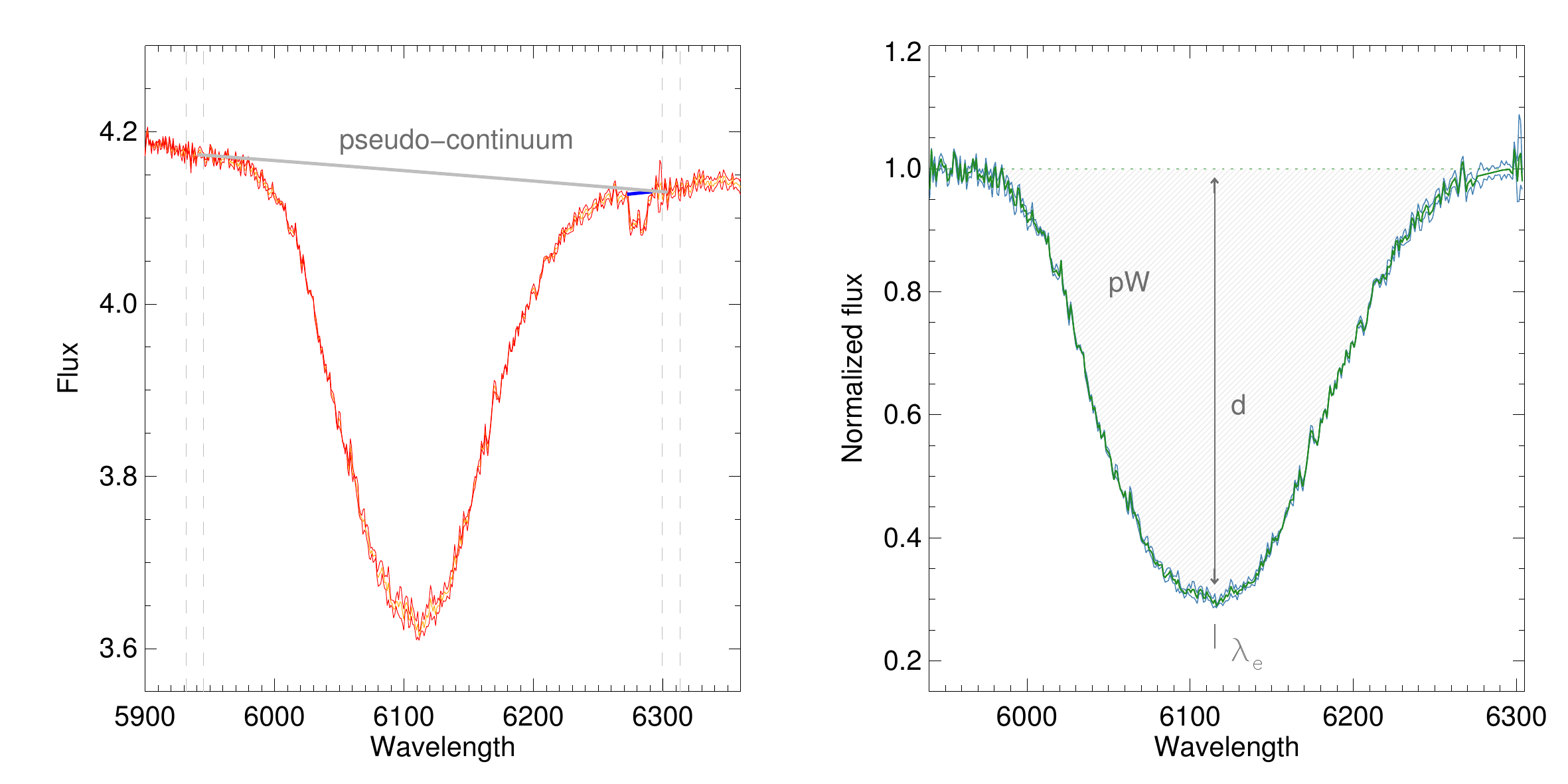}   
\caption{Description of the parameters measured in the spectra presented in this work. After determining the pseudo-continuum of the feature (as described in the text) the minimum of the feature ($\lambda_e$), its depth ($d$), and the pseudo-equivalent width ($pW$) are measured. From $\lambda_e$ and using Equation \ref{eq:vel} the expansion velocity is calculated. }
\label{fig:spe}
\end{figure}

For the most prominent features listed above we measured their expansion velocity ($v$) and pseudo-equivalent width ($pW$) when the data allowed. The depth of the feature ($d$) has been also measured for the Si II $\lambda$5972 and Si II $\lambda$6355 absorptions (See Figure \ref{fig:spe} for a representation of these three parameters).
Their measurement has been performed as follows.
First, two adjacent continuum regions 15 \AA~wide around the feature were selected.
These were used to perform a bootstrapping method using the 225 different combinations of 2 points, one from the blue and one from the red regions. For each repetition, a segment representing the pseudo-continuum was drawn and used to normalize the observed spectrum. 
In the normalized spectrum, a Gaussian fit was performed to determine the wavelength ($\lambda_e$) at which the minimum of the feature fell.
For each of the 225 repetitions, we store $\lambda_e$, the depth $d$ from the pseudo-continuum to the normalized spectrum at that wavelength, and the integral of the feature on the normalized spectrum ($pW$) from the pseudo-continuum.
Moreover, we kept all individual uncertainties coming from the instrumental flux errors.
Finally, we averaged the 225 measurements and uncertainties of $\lambda_e$, $d$, and $pW$, and used the standard deviation of their distributions as a systematic uncertainty, which has been added in quadrature to the instrumental error.
Velocities are derived from the shift of the average minimum of the feature with respect the expected rest-frame wavelength, via the relativistic Doppler formula,
\begin{equation}\label{eq:vel}
v = c  \frac{\left[(\Delta \lambda/\lambda_0)+1\right]^2-1}{\left[(\Delta \lambda/\lambda_0)+1\right]^2+1},
\end{equation}
where $\lambda_0$ is the rest-frame wavelength of the corresponding feature, and $\Delta \lambda$ is the difference between the measured wavelength $\lambda_e$ and $\lambda_0$.
In Tables \ref{tab:vel} and  \ref{tab:pew} we present the resulting measurements for $v$ and $pW$, respectively.

\begin{table*}\scriptsize
\caption{Measured velocities (in units of 10$^3$ km s$^{-1}$). Epoch referenced to the $B$ band maximum brightness.}
\label{tab:vel}
\centering        
\begin{tabular}{cccccccccc}
\hline\hline   
Epoch &Ca II H\&K&Si II 4130\AA&Mg II&Fe II   &\multicolumn{2}{c}{S II W}&Si II 5972\AA&Si II 6355\AA&Ca II trip\\
\hline
-9.6  & $-$     & $-$     & $-$     & $-$     & $-$     & $-$     & $-$     &13.6(0.1)& $-$     \\
-8.5  & $-$     &11.8(1.2)&14.3(0.2)&18.9(0.2)&11.0(0.2)&11.3(0.1)&11.5(0.1)&13.0(0.1)&14.2(0.1)\\
-7.4  &21.9(0.8)&11.0(0.2)&14.0(0.1)&18.7(0.1)&10.8(0.1)&11.1(0.1)&11.0(0.1)&12.7(0.1)&13.9(0.8)\\
-6.5  &20.4(0.9)&10.7(0.3)&14.2(0.2)&18.8(0.1)&10.7(0.1)&10.7(0.1)&10.8(0.1)&12.5(0.1)&14.0(0.3)\\
-5.4  &18.7(0.4)&10.5(0.7)&14.0(0.1)&18.4(0.1)&10.4(0.1)&10.6(0.1)&10.6(0.1)&12.2(0.1)&13.4(0.3)\\
-4.6  &18.2(0.1)&10.1(0.4)&14.0(0.3)&18.1(0.1)&10.2(0.1)&10.2(0.1)&10.6(0.1)&12.1(0.1)&13.7(0.2)\\
+2.3  & $-$     & $-$     & $-$     & $-$     & 9.9(0.1)& 9.5(0.1)&10.6(0.2)&11.5(0.1)& $-$     \\
+3.3  &16.3(0.6)& 9.6(0.1)&11.8(0.1)& $-$     & $-$     & $-$     &10.6(0.1)&11.4(0.1)& $-$     \\
+4.2  &15.3(0.2)& 9.4(0.1)&10.6(0.1)& $-$     & $-$     & $-$     &10.7(0.2)&11.2(0.1)& $-$     \\
+10.5 & $-$     & $-$     & $-$     & $-$     & 9.0(0.1)& 9.1(0.1)&12.1(0.1)&11.0(0.1)&11.9(0.1)\\
+12.4 & $-$     & $-$     & $-$     &10.0(0.1)& 8.3(0.2)& 8.7(0.2)&12.2(0.1)&10.8(0.1)& $-$     \\
+17.4 & $-$     & $-$     & $-$     & $-$     & $-$     & $-$     & $-$     &10.5(0.1)&12.0(0.1)\\
+18.3 & $-$     & $-$     & $-$     & $-$     & $-$     & $-$     & $-$     &10.3(0.1)&12.0(0.1)\\
+19.5 & $-$     & $-$     & $-$     & $-$     & $-$     & $-$     & $-$     &10.3(0.1)&12.0(0.1)\\
+22.5 & $-$     & $-$     & $-$     & $-$     & $-$     & $-$     & $-$     &10.2(0.2)&12.0(0.1)\\
+24.5 & $-$     & $-$     & $-$     & $-$     & $-$     & $-$     & $-$     &10.1(0.2)&12.1(0.1)\\
+33.4 & $-$     & $-$     & $-$     & $-$     & $-$     & $-$     & $-$     &10.2(0.2)&12.1(0.1)\\
+34.4 & $-$     & $-$     & $-$     & $-$     & $-$     & $-$     & $-$     &10.5(0.3)&12.0(0.1)\\
+39.1 & $-$     & $-$     & $-$     & 3.6(0.1)& $-$     & $-$     & $-$     &10.4(0.3)&11.8(0.1)\\
+44.4 & $-$     & $-$     & $-$     & 4.5(1.2)& $-$     & $-$     & $-$     &10.9(0.3)&12.0(0.1)\\
+66.2 & $-$     & $-$     & $-$     & $-$     & $-$     & $-$     & $-$     & 9.9(0.6)&12.2(0.1)\\
+96.2 & $-$     & $-$     & $-$     & $-$     & $-$     & $-$     & $-$     & 7.8(0.5)&12.0(0.1)\\
+97.2 &10.0(0.1)& $-$     & 2.7(0.2)& 6.1(0.7)& $-$     & $-$     & $-$     & 8.1(0.3)& $-$     \\
+99.1 &10.3(0.3)& $-$     & 2.7(0.6)& 5.8(0.2)& $-$     & $-$     & $-$     & $-$     & $-$     \\
+137.2& 9.9(0.5)& $-$     & 0.9(0.8)& $-$     & $-$     & $-$     & $-$     & 7.6(1.1)&11.9(0.1)\\
+157.2& $-$     & $-$     & $-$     & $-$     & $-$     & $-$     & $-$     & 8.2(1.6)&12.1(0.2)\\
+212.5& $-$     & $-$     & $-$     & $-$     & $-$     & $-$     & $-$     & 7.3(1.9)&11.5(0.6)\\
\hline
\end{tabular}
\end{table*}

The reddening modifies the shape of the spectrum reducing the flux more strongly in bluer wavelengths and producing an effect on spectral features. Although the minima of the spectral features is not going to be strongly affected (thus the $v$), both the slope of the pseudo-continuum and its deepness would be affected.
\cite{2011A&A...526A.119N} and \cite{2007A&A...470..411G} studied the effect of reddening in the uncertainties of $pW$ measurement and concluded that for values of $E(B-V)$ lower than 0.3 the difference is lower than 5\%, but for more extinguished SN it could be important, which is the case of SN 2014J.
To avoid a systematic error from reddening, we de?redden all spectra as described above before measuring line parameters.

\subsection{Spectral diagnostic diagrams}

Several spectral indicators have been used in the literature to study the properties of SN Ia and interpret their heterogeneities.

\cite{1995ApJ...455L.147N} defined the fractional depth of the Si II $\lambda$5972 trough the Si II $\lambda$6355 absorptions in the near-maximum light spectrum, $R({\rm Si~II})=\frac{d_{5972}}{d_{6355}}$, and showed that it correlates well with the absolute magnitude at peak, which in turn correlates with the brightness decline rate through the \cite{1993ApJ...413L.105P} relation: the more luminous the SN Ia, the lower the $R({\rm Si~II})$ at maximum light and the slower the brightness decline.
Both photometric and spectroscopic heterogeneities are attributed to differences in the effective temperature, which depend on total amount of $^{56}$Ni produced in the explosion and the kinetic energy.
The former can be estimated from the brightness decay in the bolometric light-curve, and the latter from the expansion velocity of the ejecta. 
However, the lack of correlation between the $R({\rm Si~II})$ and the photospheric velocity (deduced from Si II $\lambda$6355 blueshift) 
pointed out that only one-parameter cannot account for the SN Ia spectroscopic diversity \citep{2000ApJ...543L..49H}.
In fact, the current widely spread SN Ia photometric standardization needs two parameters (stretch and color, \citealt{2007A&A...466...11G}; \citealt{2008ApJ...681..482C}; \citealt{2007ApJ...659..122J}) to reduce the scatter on the peak magnitudes, and even a third parameter accounting for the environment has been proposed \citep{2010ApJ...722..566L, 2010MNRAS.406..782S, 2014A&A...568A..22B, 2015arXiv151105348M}.

\begin{figure*}
\centering
\includegraphics[trim=1.2cm 0.4cm 0.6cm 0.8cm, clip=true,width=0.475\linewidth]{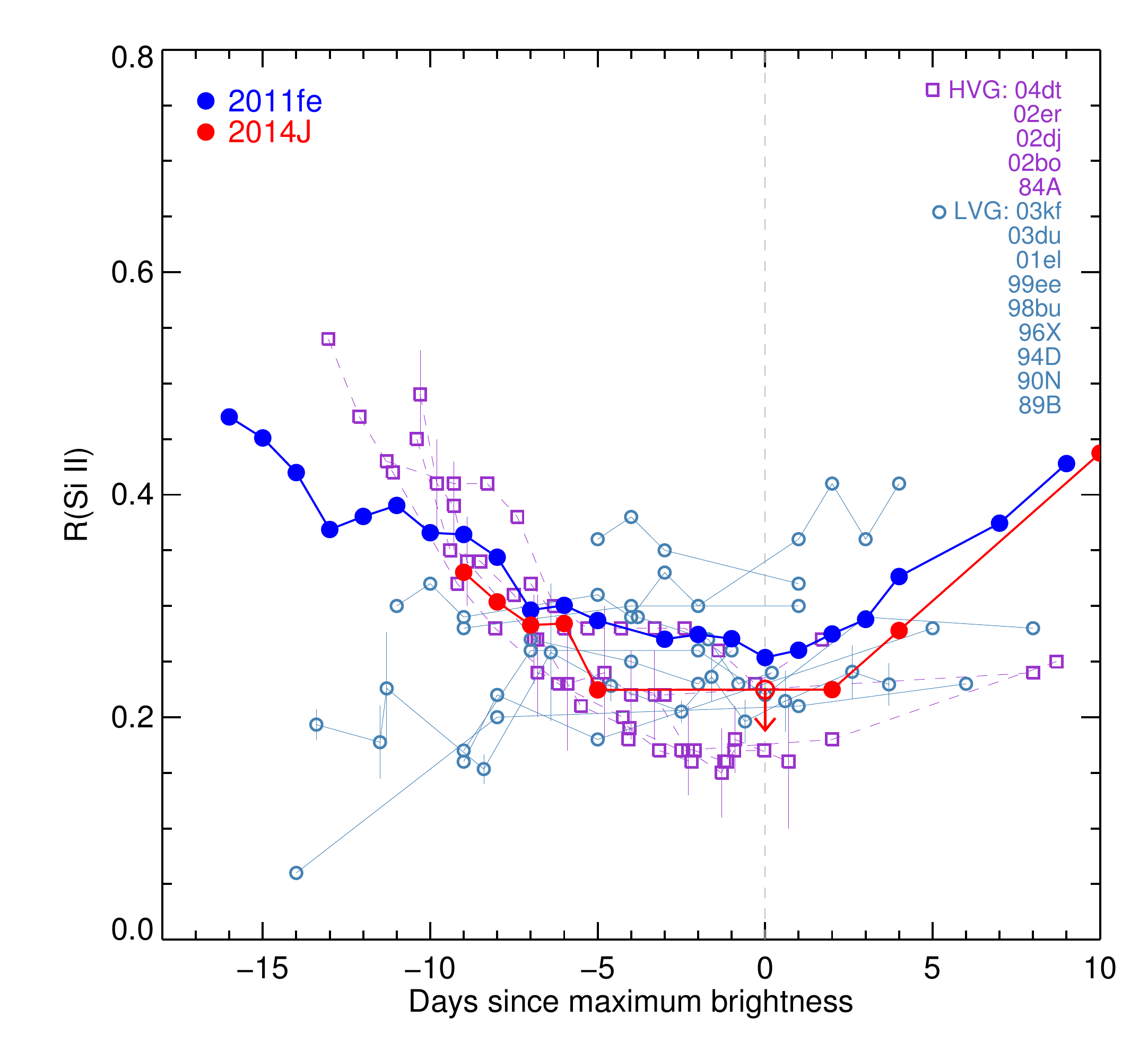}   
\includegraphics[trim=1.1cm 0.4cm 0.8cm 0.8cm, clip=true,width=0.519\linewidth]{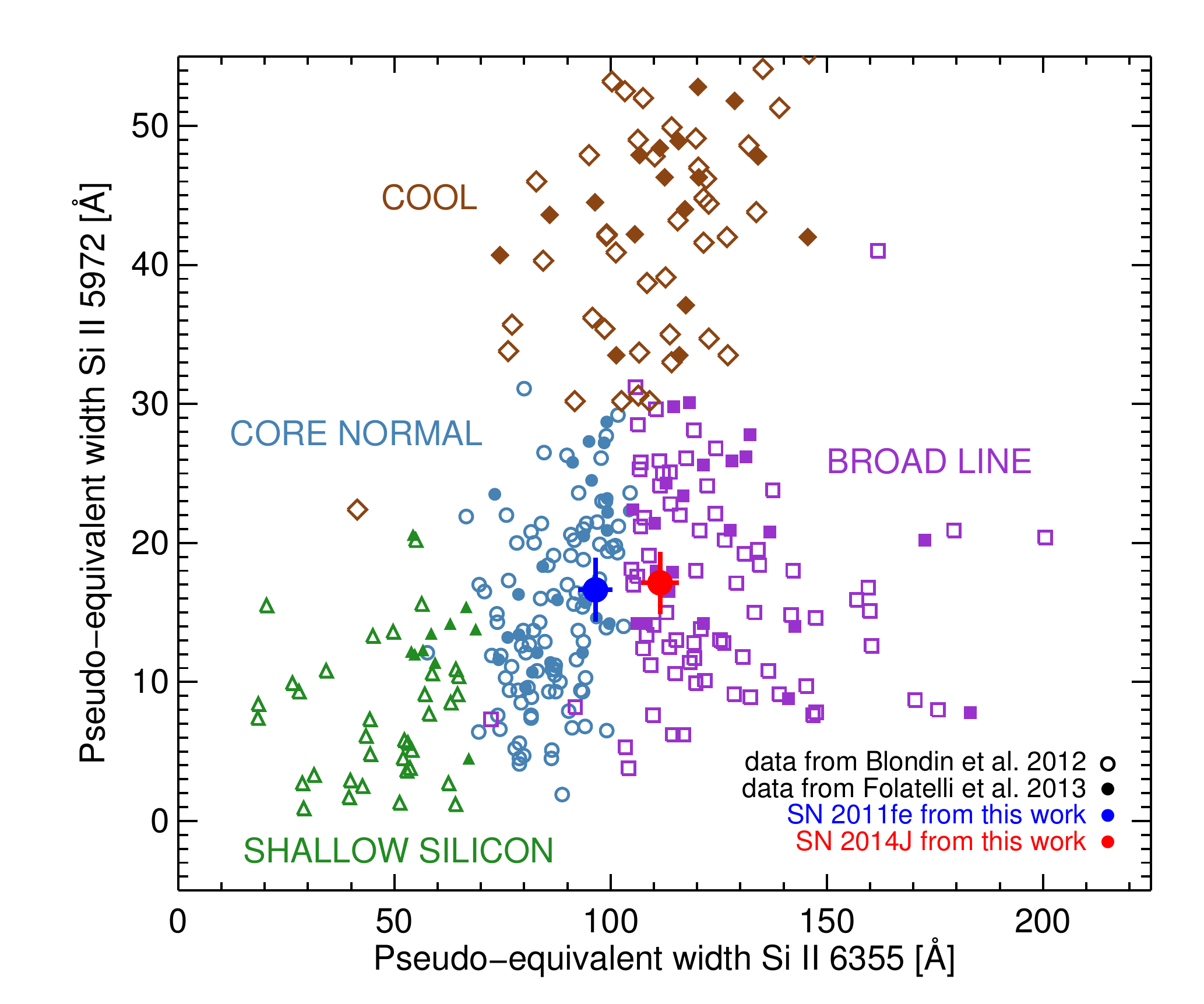}   
\caption{Left: Diagram from \protect\cite{2005ApJ...623.1011B} where the evolution of the $R({\rm Si~II})$ parameter is compared for {\sc HVG} (blue) and {\sc LVG} (purple) SN groups. Original data from \protect\cite{2009ApJ...695..135A} are shown, and SN 2011fe and SN 2014J are overplotted. Right: Diagram from \protect\cite{2006PASP..118..560B} where the pseudo-equivalent widths of the two SI II features at 5972 and at 6355 are compared. Data from the Center for Astrophysics Supernova Program (CfA, \protect\citealp{2012AJ....143..126B}, open symbols) and from the Carnegie Supernova Project (CSP, \protect\citealp{2013ApJ...773...53F}, filled symbols) populate the diagram, and our measurements situate SN 2011fe (blue) and SN 2014J (red) on top.}
\label{fig:alt}
\end{figure*}

BE05 studied the evolution of the $R({\rm Si~II})$ ratio and proposed that SN Ia could be well separated in two groups depending on their value at maximum:
(i) High velocity gradient ({\sc HVG}) SN Ia, with high $R({\rm Si~II})$ values right after explosion and which decreases monotonically to $R({\rm Si~II})\lesssim$ 0.2 around the epoch of maximum brightness; and
(ii) Low Velocity Gradient ({\sc LVG}) SN Ia show either no evolution or increasing $R({\rm Si~II})$ values from explosion up to the peak brightness, when they tend to have higher $R({\rm Si~II})$ values than {\sc HVG} SN Ia.
Additionally, by using a cluster analysis with other photometric and spectral properties, they defined a third group of underluminous SN Ia ({\sc FAINT}) which actually have $R({\rm Si~II})>$ 0.4 at the epoch of maximum brightness.
This separation was even clearer when comparing the Si II $\lambda$6355 velocity gradient ($\dot{v}_{10}$, from the epoch of maximum brightness up to 10 days post-maximum) with the $\Delta$m$_{15}$ decline rate parameter. While {\sc FAINT} SN Ia show higher $\Delta$m$_{15}$ values than the other two groups, {\sc LVG} and {\sc HVG} are disentangled by their velocity gradient ($\langle\dot{v}_{10}\rangle_{\rm HVG}$ = 97 $\pm$ 16, and $\langle\dot{v}_{10}\rangle_{\rm LVG}$ = 37 $\pm$ 18 km s$^{-1}$ d$^{-1}$ in the original BE05 sample).

Finally, this separation was also interpreted as differences in the mechanism responsible for the explosion (while {\sc HVG} SN Ia could be produced by delayed-detonations, {\sc LVG} SN would be the result of deflagrations), different heavy element mixing in the structure of the WD (more efficient for {\sc HVG} SN Ia, and less in {\sc LVG} SN Ia), or differences in the viewing angle assuming asymmetric explosions (two SNe Ia that are physically identical in three dimensions could be classified differently just because they are seen from different directions; \citealt{2010ApJ...708.1703M}).
There is a remarkable continuity in the $R({\rm Si~II})$ parameter at maximum light, enabling the presence of extreme, peculiar, and intermediate class objects.
 
BR06 proposed a different classification by constructing a diagram from the pseudo-equivalent widths of the same Si II features used by \cite{1995ApJ...455L.147N} measured at maximum light.
Although the spectral features used in BE05 and BR06 diagrams are the same, the depth of the feature used in the BE05 and the shape, width, and strength with respect the adjacent continuum used in the BR06 are not showing exactly the same information.
Based on the position of the SN Ia in that diagram and on the actual appearance of the Si II $\lambda$6355 feature, BR06 distinguished four different groups:
(i) Shallow-silicon SN Ia ({\sc SS}) show small $pW$ values in the two features ($\lesssim$ 70~\AA~for Si II $\lambda$6355 and $\lesssim$ 25~\AA~for Si II $\lambda$5972);
(ii) Core-normal SN Ia ({\sc CN}) have similar $pW$ and shape of the $\lambda$6355 absorption, and higher $pW$  Si II $\lambda$5972 values than {\it shallow silicon} SN Ia (up to 105~\AA). Differences were due to lower temperatures than the former; 
(iii) Broad-line SN Ia ({\sc BL}) have even higher Si II $\lambda$6355 $pW$ values ($\gtrsim$ 105~\AA), and show broader and deeper absorptions than those of the {\it core-normal} SN Ia;
and (iv) Cool SN Ia ({\sc CL}), which have higher $pW$ for both features, specially higher Si II $\lambda$5972 $pW$ compared to the other groups ($\gtrsim$ 30~\AA).
As in the BE05 diagram, the intermediate regions between groups are populated with intermediate objects, showing a sequential continuity in the spectral properties for SN Ia.

There is a clear correspondence between both diagrams detailed above (see e.g. \citealt{2014Ap&SS.351....1P}).
BR06 {\sc CL} SN Ia correspond to the BE05 {\sc FAINT} SN Ia, which is to be expected, since both temperature and luminosity are controlled mainly by the $^{56}$Ni mass. 
BR06 {\sc BL} SN Ia correspond to the BE05 {\sc HVG} SN Ia. This also makes sense, because broad Si II $\lambda$6355 absorption requires high Si II optical depth over a substantial velocity range, which makes it possible for the absorption minimum to shift appreciably with time. {\sc BL} SN Ia may have thicker silicon layers than the core-normal SN Ia. 
Both BR06 {\sc CN} and {\sc SS} SN Ia correspond to the BE05 {\sc LVG} SN Ia. The lower velocity range over which Si II has a high optical depth permits only a smaller shift in the absorption minimum with time.

Lately, WA09 proposed a different diagram between the pseudo-equivalent width of the Si II $\lambda$6355 absorption and the velocity of the same feature near $B$ band maximum light, instead of the $pW$ Si II $\lambda$5972 absorption used in BR06 diagram. 
They distinguished between {\sc normal} SN Ia and high-velocity ({\sc HV}), which showed a linear trend in the direction of higher velocity for higher $pW$, and also defined two other groups: {\it underluminous} SN Ia (which corresponds to 91bg-like SN Ia) with lower $pW$ and lower velocities at maximum, and {\it overluminous SN Ia} (corresponding to 91T-like SN Ia) with lower velocities but similar $pW$ than {\sc normal} SN Ia.
This description has one-to-one correspondence with the BR06 groups.

This scheme summarizes the spectroscopic diversity discovered so far: a decreasing temperature sequence from 1991T-like ({\sc SS}) to normal ({\sc CN}) and 1991bg-like SN Ia ({\sc CL}), plus the high-velocity SN ({\sc BL}) as a branch from the normal SN Ia group.

\begin{figure}
\centering
\includegraphics[trim=1.1cm 0.3cm 0.5cm 1.2cm, clip=true,width=\linewidth]{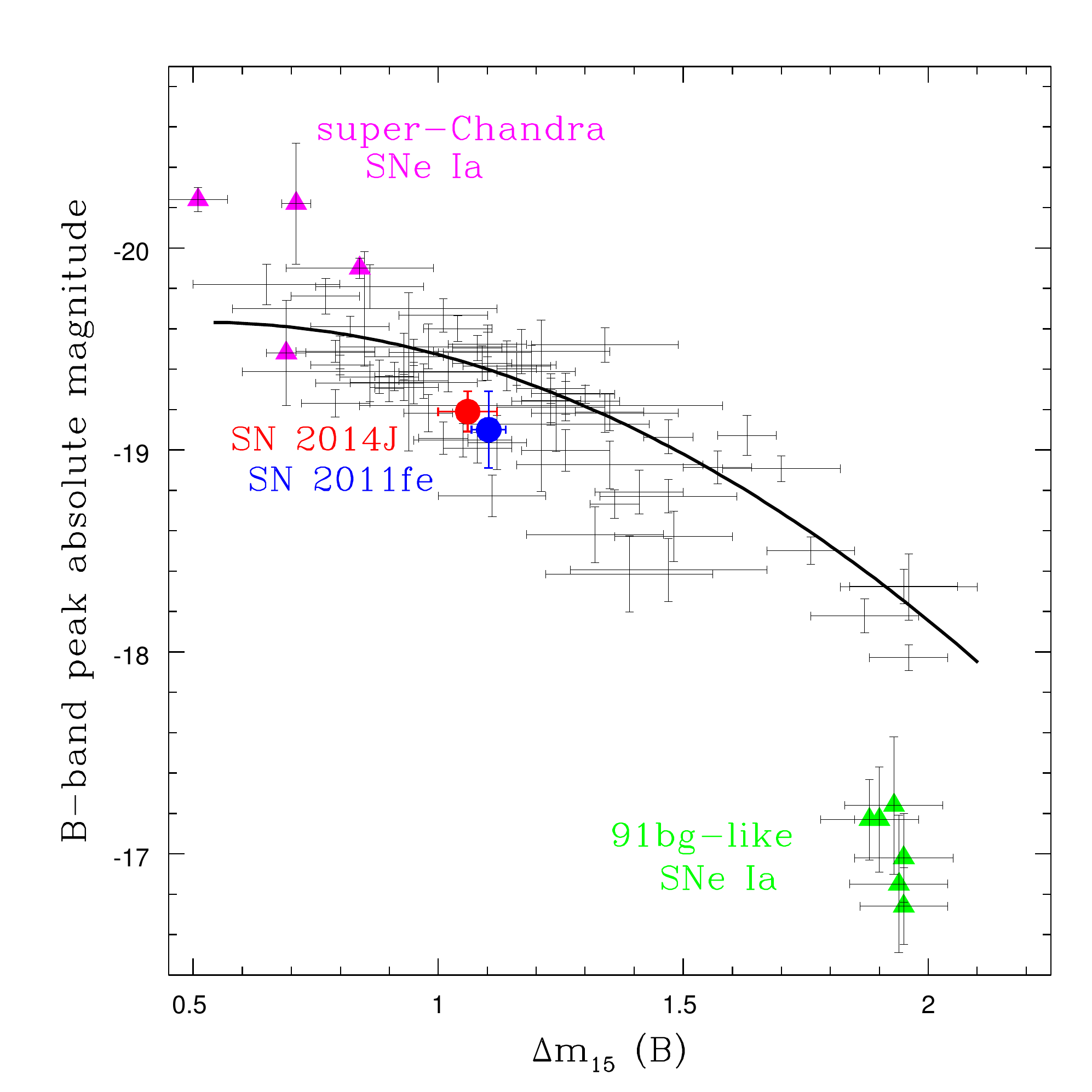}   
\caption{Relation between the $B$-band peak absolute magnitude and brightness decline in the $B$-band fifteen days after maximum ($\Delta$m$_{15}$(B), \citealt{1993ApJ...413L.105P,1538-3881-118-4-1766}). The solid line represents the Phillips relation, that is followed by all {\it normal} SN Ia. Both SN 2011fe and SN 2014J are located on the bulk of normal objects. Although low-luminosity events ({\it 91bg-like}) are well below the relation, the overluminous SN Ia ({\it super-Chandra}) may tentatively follow the solid line.}
\label{fig:phil}
\end{figure}

\section{Results and discussion} \label{Section4}

\subsection{SN 2014J in BE05, BR06 and WA09 diagrams}

In Figure \ref{fig:alt} we show the two diagnostic diagrams by BE05 and BR06.
In this and in the following Figures, SN 2011fe has also been shown as a reference for comparisons. It is also a nearby object which shows few signs of extinction ($E(B-V)\sim0.03$, accounting for both MW and host galaxy reddening, \citealt{2014MNRAS.439.1959M}).
Spectra has been downloaded from WISeREP and the same spectral parameters have been measured in an homogeneous way for this work.\footnote{Although, as stated in the text, SN 2011fe spectra have been download from WISeREP, the actual sources that published the data used in this paper are \cite{2012ApJ...752L..26P}, \cite{2013A&A...554A..27P}, \cite{2014MNRAS.439.1959M}, and \cite{2014MNRAS.444.3258M}.} 

The BE05 diagram have been filled with the original objects from BE05 and \cite{2009ApJ...695..135A}. {\sc HVG} SN Ia (in purple) show higher values of $R({\rm Si~II})$ in the pre-maximum spectra which decrease near maximum light, while for {\sc LVG} SN Ia the $R({\rm Si~II})$ parameter evolves increasing monotonically or show no evolution at all.
Compared to Si II $\lambda$6355, the Si II $\lambda$5972 absorption is produced by a transition with a higher excitation energy, so $R({\rm Si~II})$ should increase in strength for higher temperatures \citep{1995ApJ...455L.147N}.
This has been interpreted by BE05 as {\sc HVG} SN having cooler temperatures at the line-forming regions that increase approaching maximum, while {\sc LVG} SN, on the other hand, have high temperatures already well before maximum.
\cite{1995ApJ...455L.147N} suggested the reason: at lower temperatures the blanketing from Fe II and Co II increase the apparent strength of Si II $\lambda$5972,
 at higher temperatures Fe III and Co III wash out the feature.
Moreover, \cite{1995ApJ...455L.147N} showed a correlation between $R({\rm Si~II})$ at maximum and the peak absolute magnitude, in the direction of brighter SN having lower $R({\rm Si~II})$ values.

$R({\rm Si~II})$ measurements for both SN 2014J and SN 2011fe are listed in Table \ref{tab:rsi}.
SN 2011fe has $R({\rm Si~II})$ values higher than SN 2014J during the whole period shown in the left panel of Figure \ref{fig:alt}. In the pre-maximum phase this can be interpreted as SN 2011fe having lower photospheric temperature due to more line blanketing by Fe and Co, and in the near maximum phase peaking at fainter absolute magnitude. 
Although the reported values in the literature point to similar or even slightly brighter $M_B$ for SN 2011fe than for SN 2014J (14J: -19.26 $\pm$ 0.26 mag \citealt{2014ApJ...795L...4K}, -19.19 $\pm$ 0.10 \citealt{2015ApJ...798...39M}; 11fe: -19.45 $\pm$ 0.08 \citealt{2011arXiv1112.0439T}, -19.21 $\pm$ 0.15 \citealt{2012JAVSO..40..872R}) the $\Delta$m$_{15}$ values for SN 2011fe reported are slightly higher than for 2014J.

According to the relation between the $B$-band peak absolute magnitude and the $\Delta$m$_{15}$(B) parameter, which accounts for the brightness decline in the $B$-band fifteen days after maximum \citep{1993ApJ...413L.105P,1538-3881-118-4-1766}, these two SN lay perfectly on top of the bulk relation of {\it normal} SN Ia, well far from the subluminous SN Ia-91bg, and the overluminous super-Chandra SN Ia, as shown in Figure \ref{fig:phil}.
This supports the classification of SN 2014J as a {\it normal} SN Ia from a photometric point of view, allowing its use for cosmological analyses.

\begin{figure}
\centering
\includegraphics[trim=0.6cm 0.2cm 0.6cm 0.6cm, clip=true,width=\linewidth]{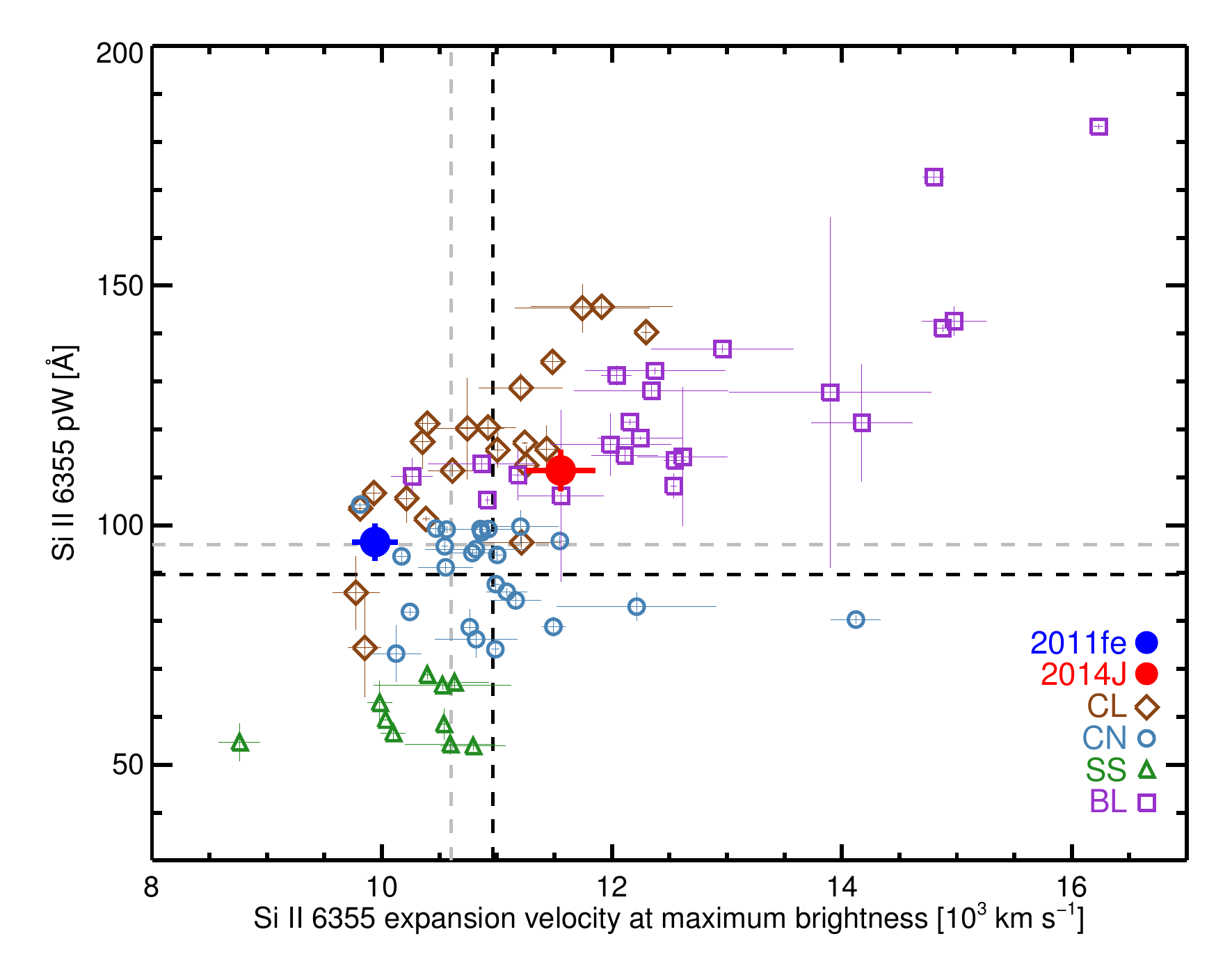}   
\caption{Diagram from \protect\cite{2009ApJ...699L.139W}  comparing the pseudo-equivalent width of the Si II $\lambda$6355 feature at $B$ band maximum light with the velocity of the same feature at the same epoch. The diagram is populated with the SN Ia of the CSP taken from \protect\citealp{2013ApJ...773...53F}. SN 2011fe are located closer to the 'normal' SN Ia group, while SN 2014J is in between the 'HV' and the 'normal' groups. Horizontal and vertical grey lines represent the average $pW$ and $v$ of the 'normal' group reported in \protect\cite{2009ApJ...699L.139W}, and the black lines the average values for the CSP sample.}
\label{fig:wang}
\end{figure}

Due to the lack of very early $R({\rm Si~II})$ measurements for SN 2014J no definitive classification can be done using BE05 diagram, although it seems to tentatively follow other {\sc LVG} SN Ia behavior, and definitely have lower $R({\rm Si~II})$ value at maximum than SN 2011fe. However, the near maximum light $pW$ measurements allowed to position 2014J in the BR06 diagram (See right panel in Figure \ref{fig:alt}). 
The BR06 diagram has been populated with data from the Center for Astrophysics Supernova Program (CfA, \citealp{2012AJ....143..126B}) and from the Carnegie Supernova Project (CSP, \citealp{2013ApJ...773...53F}) to define the regions covered by each of the four groups.
Although SN 2011fe falls within the {\sc CN} sector (corresponding to {\sc LVG} in BE05 diagram)  and SN 2014J are within the {\sc BL} region (which corresponds to {\sc HVG} in BE05), both are located close to the border defined between the two groups. 
The resulting BR06 classification is in agreement with what we found in BE05: while both objects tend to follow the behaviors of their corresponding groups, they both seem to be extreme objects of their classes and similar to each other.

\begin{figure}
\centering
\includegraphics[trim=1.2cm 0.45cm 0.6cm 0.95cm, clip=true,width=\linewidth]{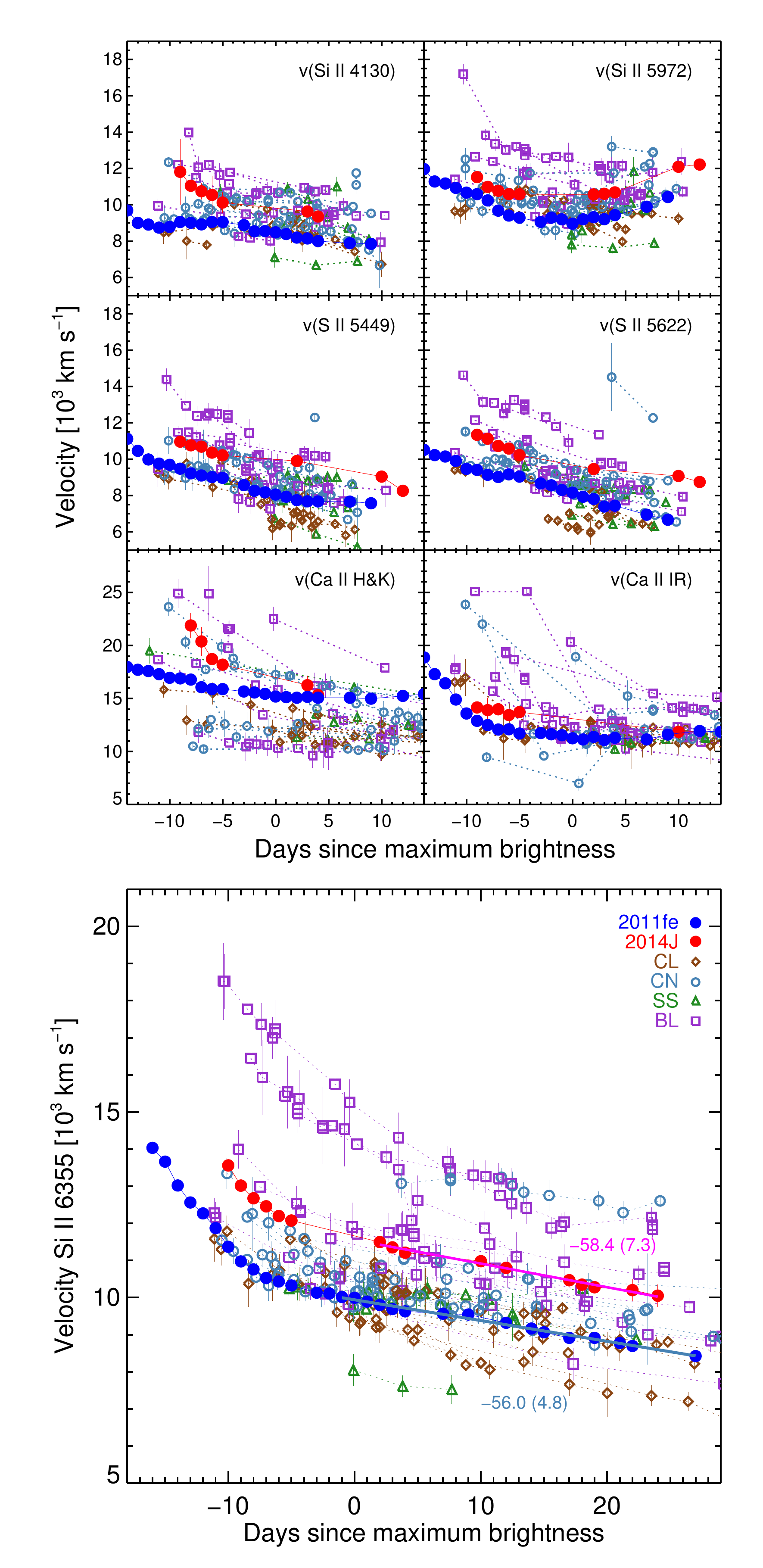}   
\caption{Velocity evolution of the main features in SN Ia spectra. Si II, S II, and Ca pairs are compared on top, and below the Si II $\lambda$6355 absorption velocity evolution are presented. SN 2011fe (in blue) and SN 2014J (in red) are shown on top of the CSP sample. In the bottom panel two linear fits have performed to measure $\dot{v}_{10}$ for both SN 2011fe and 2014J.}
\label{fig:gas2}
\end{figure}

\begin{figure}
\centering
\includegraphics[trim=0.0cm 0.6cm 0.7cm 0.9cm, clip=true,width=\linewidth]{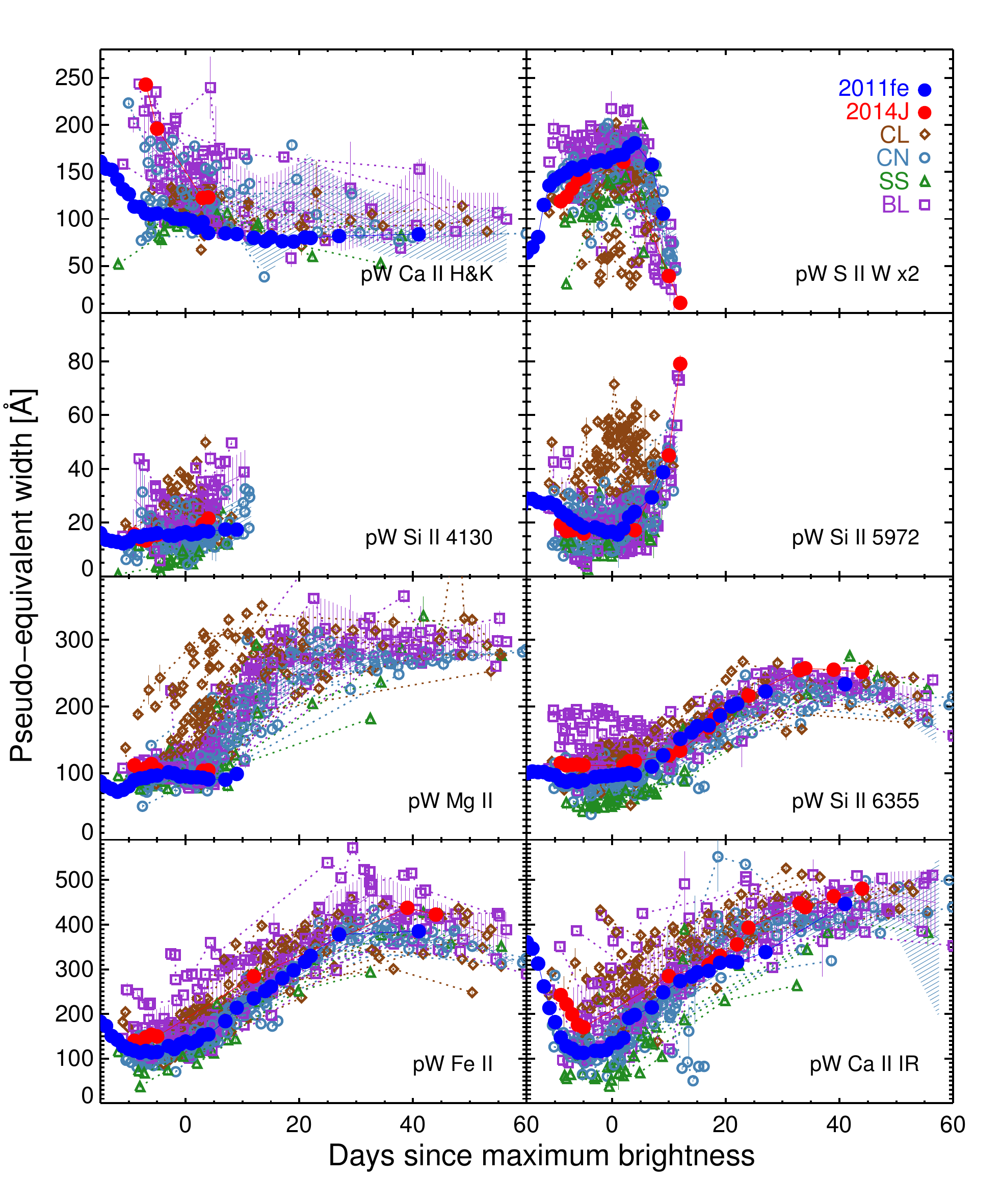}   
\caption{Pseudo-equivalent width evolution of the SN 2014J (in red) and SN 2011fe (in blue). The evolution of the pW in each feature for the four groups defined by BR06 obtained by measuring the pW from the CSP SN Ia sample are also shown as a reference. The shaded bands show the average and 1$\sigma$ dispersion of {\sc CN} and {\sc BL} SN Ia.}
\label{fig:gas1}
\end{figure}

As a cross-check for the classification provided by these diagrams, in Figure \ref{fig:wang} we show the corresponding diagram presented by WA09. 
Since SN 2014J also shows higher velocities than SN 2011fe, it is also positioned within the {\sc BL} group, but in the border with the {\sc CN} region. 

\subsection{Velocity and pseudo-EW evolution}

Figure \ref{fig:gas2} shows the evolution of the expansion velocities of Si II, S II and Ca II features.
In the background, we plotted in each panel the evolution of the velocities of the CSP SN Ia presented in \cite{2013ApJ...773...53F} colored by the spectroscopic group defined in BR06 diagram.
The overall expansion velocities of all features are higher for SN 2014J with respect SN 2011fe.
We note that in all panels, while SN 2011fe is on top of other {\sc CN} SN Ia velocity evolution, SN 2014J is on the bottom end of the {\sc BL} group, which stresses it being a {\sc BL} SN, but very close to the {\sc CN} group, and is in agreement to what we found in the spectroscopic diagrams.
For the Si II $\lambda$6355 absorption we calculated the $\dot{v}_{10}$ for both SN. SN 2014J shows a velocity gradient of -58.4 $\pm$ 7.3 km s$^{-1}$ d$^{-1}$, and for SN 2011fe we found -56.0 $\pm$ 4.8 km s$^{-1}$ d$^{-1}$, in agreement with \cite{2013A&A...554A..27P} who found -59.6 $\pm$ 3.2 km s$^{-1}$ d$^{-1}$. Although SN 2014J's $\dot{v}_{10}$ is faster than the value found for SN 2011fe, both values are within the quoted errors.

Figure \ref{fig:gas1} shows the evolution of the pseudo-equivalent widths of the eight strongest features measured in this work.
The average evolution (and the 1$\sigma$ deviation) of the $pW$ in each feature for the four groups defined by BR06 obtained by measuring the $pW$ from the CSP SN Ia sample are also shown for reference.
In all panels the {\sc BL} and {\sc CN} strips are mostly overlapping, making difficult any association of the studied objects to any of these groups.
The differences are clearer in the evolution of the {\sc SS} and {\sc CL} groups. {\sc CL} SN Ia show higher $pW$ Mg II, and $pW$ Si II $\lambda$5972, and lower $pW$ S II W values, while {\sc SS} have lower $pW$ in all features except in the S II W absorption.

\begin{figure}
\centering
\includegraphics[trim=0.3cm 0.3cm 0.8cm 0.9cm, clip=true,width=\linewidth]{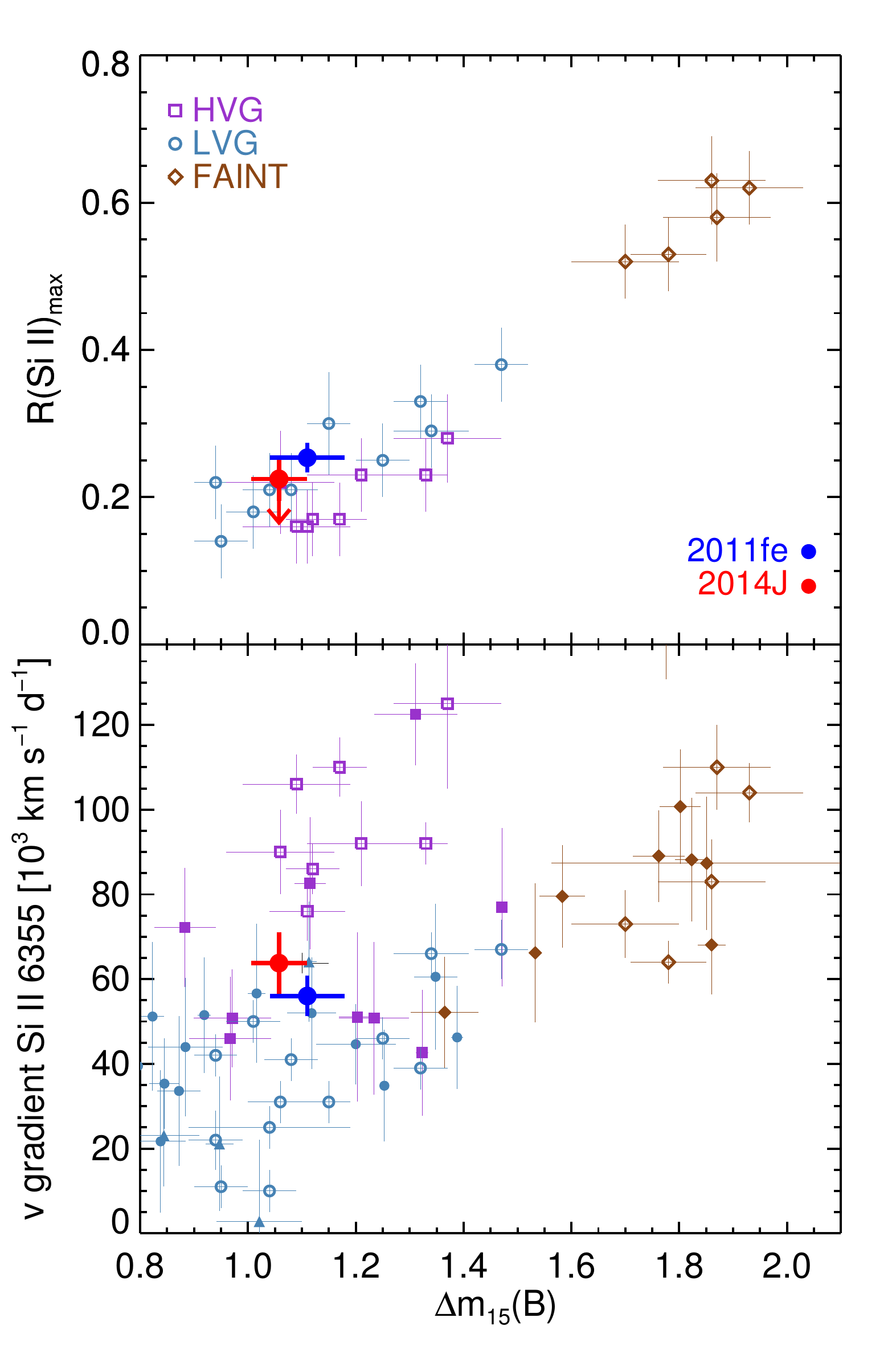}   
\caption{$R({\rm Si~II})$ and v(Si II 6355) vs $\Delta$m$_{15}$(B) diagram. Data from \protect\cite{2009ApJ...695..135A} and \protect\cite{2005ApJ...623.1011B} is shown with unfilled symbols, and data from the CSP with filled symbols. SN 2014J (in red) and SN 2011fe (in blue) are situated showing that both are intermediate objects between {\sc LVG} and {\sc HVG} groups.}
\label{fig:ben}
\end{figure}

SN 2014J seems to be more associated to the {\sc BL} class if one accounts for the behavior of the Ca II H\&K $pW$. The early higher values are characteristic of this group compared to the lower values that both the {\sc CN} group and SN 2011fe show.
This is confirmed by the lower $pW$ values found post-maximum for SN 2014J and the {\sc BL} group in the S II W feature, and the slightly higher $pW$ values for the Si II $\lambda$4130, SI II $\lambda$5972, Fe II, and Ca II IR features.
For Mg II we only were able to measure the $pW$ up to $\sim$10 days past-maximum light, and no conclusions can be made from this early phase.
Finally, the Si II $\lambda$6355 $pW$ evolution of both SN 2011fe and SN 2014J is very similar, although the value at maximum light is higher for 14J than for 11fe.
In general, {\sc BL} covers the SN 2014J evolution, and SN 2011fe $pW$s evolution follow the {\sc CN} strip. However, they both are very similar, which underlines the proximity of the two objects within the two groups.

\subsection{Spectral properties vs. $\Delta$m$_{15}$(B)}

Figure \ref{fig:ben} in its top panel shows the existing correlation between $R({\rm Si~II})$ and $\Delta$m$_{15}$(B). Since $R({\rm Si~II})$ traces the temperature, and $\Delta$m$_{15}$ the brightness, the scatter the {\sc LVG} objects introduces in the tight correlation between {\sc HVG} and {\sc FAINT} groups, can be interpreted as a need for a different physical parameter besides temperature to explain the heterogeneity of SN Ia, as discussed in BE05.
Both SN 2011fe and 2014J follow the linear trend and, in this diagram SN 2014J is closer to the {\sc LVG} SN Ia behavior. 

We showed in Figure \ref{fig:gas2} that both SN Ia have similar velocity gradients. Their $\Delta$m$_{15}$ (1.06 $\pm$ 0.06 for 2014J, averaging the values reported by \citealt{2014CoSka..44...67T,2014MNRAS.445.4427A,2015ApJ...798...39M,2014ApJ...795L...4K}, and 1.11 $\pm$ 0.07 for SN 2011fe, from \citealt{2011arXiv1112.0439T,2012JAVSO..40..872R,2013ApJ...767..119M,2013A&A...554A..27P}) are similar as well, so in the bottom panel of Figure \ref{fig:ben}, they are positioned in the overlapping region between the {\sc LVG} and the {\sc HVG} groups.
Here we include both the BE05 objects and CSP objects. They seem to agree well with the regions defined by their classifications.
{\sc FAINT} SN Ia have large $\Delta$m$_{15}$ and $\dot{v}$ values, while the other two groups have lower $\Delta$m$_{15}$ and are separated by their  $\dot{v}$. 
Here, the expansion velocity gradient, $\dot{v}$, seems to be weakly correlated with $\Delta$m$_{15}$(B) for {\sc LVG} and {\sc FAINT} groups, while {\sc HVG} SN Ia are separate from {\sc LVG} by their larger $\dot{v}$ values.

\begin{table*}\scriptsize
\caption{Measured pseudo-equivalent widths (in \AA). Epoch referenced to the $B$ band maximum brightness.}
\label{tab:pew}
\centering        
\begin{tabular}{ccccccccc}
\hline\hline   
Epoch &Ca II H\&K &Si II 4130\AA&Mg II  &Fe II      &S II W   &Si II 5972\AA&Si II 6355\AA&Ca II trip\\
\hline
-9.6  &$-$        &$-$      &$-$        &$-$        &$-$      &$-$      &$-$       &$-$          \\
-8.5  &$-$        &15.6(3.1)&111.7(6.2) &139.8(6.7) &59.4(4.5)&19.3(1.5)&115.2(2.9)&242.2(2.7)\\
-7.4  &$-$        &13.9(1.5)&102.3(1.2) &136.0(1.6) &61.7(2.8)&16.8(1.2)&111.6(2.4)&221.7(4.2)\\
-6.5  &242.8(5.8) &13.3(2.6)&107.3(3.6) &147.3(3.5) &66.1(3.8)&17.2(1.3)&112.6(3.1)&198.9(2.8)\\
-5.4  &$-$        &15.2(6.7)&113.5(4.5) &153.2(3.4) &71.0(3.2)&17.5(1.1)&113.0(4.6)&175.6(3.3)\\
-4.6  &195.9(8.0) &15.5(3.4)&105.3(4.9) &149.5(5.5) &72.1(3.8)&15.9(1.9)&112.0(2.3)&169.9(3.4)\\
+2.3  &$-$        &$-$      &$-$        &$-$        &80.3(3.5)&17.6(2.6)&111.2(3.7)&$-$       \\
+3.3  &122.3(6.0) &19.4(1.0)&103.4(1.9) &$-$        &$-$      &$-$      &117.8(3.8)&$-$       \\
+4.2  &123.0(3.2) &21.5(1.4)&104.3(2.9) &$-$        &$-$      &17.2(1.6)&118.3(4.6)&$-$       \\
+10.5 &$-$        &$-$      &$-$        &$-$        &19.6(1.5)&45.0(2.7)&126.6(3.0)&284.8(2.9)\\
+12.4 &$-$        &$-$      &$-$        &284.2(7.6) & 5.4(3.1)&79.1(3.1)&133.6(2.3)&$-$       \\
+17.4 &$-$        &$-$      &$-$        &$-$        &$-$      &$-$      &168.6(5.3)&309.8(2.6)\\
+18.3 &$-$        &$-$      &$-$        &$-$        &$-$      &$-$      &181.1(5.5)&318.0(2.4)\\
+19.5 &$-$        &$-$      &$-$        &$-$        &$-$      &$-$      &187.9(6.3)&330.1(2.4)\\
+22.5 &$-$        &$-$      &$-$        &$-$        &$-$      &$-$      &203.2(6.6)&355.8(1.7)\\
+24.5 &$-$        &$-$      &$-$        &$-$        &$-$      &$-$      &216.1(5.8)&393.2(4.8)\\
+33.4 &$-$        &$-$      &$-$        &$-$        &$-$      &$-$      &254.2(5.9)&447.7(3.3)\\
+34.4 &$-$        &$-$      &$-$        &$-$        &$-$      &$-$      &257.2(5.9)&439.2(5.6)\\
+39.1 &$-$        &$-$      &$-$        &437.6(11.9)&$-$      &$-$      &255.1(5.3)&463.4(3.5)\\
+44.4 &$-$        &$-$      &$-$        &422.4(4.7) &$-$      &$-$      &251.7(5.8)&480.6(4.2)\\
+66.2 &$-$        &$-$      &$-$        &$-$        &$-$      &$-$      &245.8(8.9)&525.4(5.0)\\
+96.2 &$-$        &$-$      &$-$        &$-$        &$-$      &$-$      &240.0(9.1)&553.8(6.2)\\
+97.2 &137.6(9.4) &$-$      &289.2(10.3)&252.0(8.4) &$-$      &$-$      &245.8(3.9)&$-$       \\
+99.1 &165.1(16.0)&$-$      &317.3(12.2)&282.7(7.2) &$-$      &$-$      &$-$       &$-$       \\
+137.2&152.3(7.0) &$-$      &279.3(5.8) &$-$        &$-$      &$-$      &254.0(5.2)&494.9(4.7)\\
+157.2&$-$        &$-$      &$-$        &$-$        &$-$      &$-$      &254.6(6.2)&453.6(10.5)\\
+212.5&$-$        &$-$      &$-$        &$-$        &$-$      &$-$      &271.0(8.2)&412.5(34.8)\\
\hline
\end{tabular}
\end{table*}

\begin{table}\small
\caption{$R({\rm Si~II})$ evolution for both SNe 2011fe and 2014J.}
\label{tab:rsi}
\centering        
\begin{tabular}{ccc}
\hline\hline   
Epoch   & 2011fe & 2014J\\
\hline
-16 & 0.47 & --   \\
-15 & 0.45 & --   \\
-14 & 0.42 & --   \\
-13 & 0.37 & --   \\
-12 & 0.38 & --   \\
-11 & 0.39 & --   \\
-10 & 0.37 & --   \\
-9  & 0.36 & 0.33 \\
-8  & 0.34 & 0.30 \\
-7  & 0.30 & 0.28 \\
-6  & 0.30 & 0.28 \\
-5  & 0.29 & 0.22 \\
-3  & 0.27 & --   \\
-2  & 0.27 & --   \\
-1  & 0.27 & --   \\
 0  & 0.25 & --   \\
 1  & 0.26 & --   \\
 2  & 0.27 & 0.22 \\
 3  & 0.29 & --   \\
 4  & 0.33 & 0.28 \\
 7  & 0.37 & --   \\
 9  & 0.43 & --   \\
 10 & --   & 0.44 \\
 12 & --   & 0.72 \\ 
 \hline
\end{tabular}
\end{table}

In all diagnostics, SN 2014J follows the characteristics of {\sc BL}/{\sc HVG} SN Ia, being near the boundaries between {\sc CN}/{\sc LVG} subclasses.
SN 2011fe is in the opposite situation falling in the {\sc CN}/{\sc LVG} regions in all parameter spaces, but close to the {\sc BL}/{\sc HVG} sector.
Looking at most of the parameters studied in this work, the differences between SN 2011fe and 2014J are within the reported uncertainties.
The subclasses in all diagnostics are not distinct in the sense that they are not separated by regions. However, SN Ia appear to be better described as a continuum than distinct subclasses, and SN 2011fe and 2014J being that close in the border of their subclasses is further evidence that the subclassifications are artificial.


\section{Conclusions} \label{Section5}

We present the compilation of 27 spectra observed between January 22$^{\rm nd}$ and September 1$^{\rm st}$ 2014 with the 2.5m Isaac Newton and 4.2m William Herschel Telescopes, located at the El Roque de Los Muchachos Observatory, La Palma.
Observations were taken with different instruments and technical configurations. These and the reduction from the raw data are also described.

We measured several spectral parameters, including velocities, pseudo-equivalent widths, and absorption depths, for the most prominent features, and studied their evolution with time.
With this information, and using SN 2011fe for comparison, we discussed the position of SN 2014J in \cite{2005ApJ...623.1011B}, \cite{2006PASP..118..560B}, and  \cite{2009ApJ...699L.139W} diagrams showing that SN 2014J is an intermediate object between the {\it core-normal} and {\it broad-line} SN Ia (according to \citealt{2006PASP..118..560B}), also intermediate between the high velocity gradient and low velocity gradient groups (according to \citealt{2005ApJ...623.1011B}), and in between the normal and high-velocity groups (according to \citealt{2009ApJ...699L.139W}).
The most noticeable difference between the two objects is the difference of 1,500 km s$^{-1}$ in their Si II 6355 expansion velocities around maximum light, as seen in the \cite{2009ApJ...699L.139W} diagram.
These diagnostic diagrams give more information about the state of the ejecta than the possible explosion mechanisms progenitor scenarios. In that sense, the proposed subclasses are better described by a continuous sequence than distinct groups, and this is supported by the existence of transitional SN Ia as SN 2011fe and SN 2014J.
These findings support the classification of this middle-class SN 2014J as a standard event \citep{2015ApJ...812...62C} very similar to the bulk of SN Ia useful for measuring cosmological distances.
  
\section*{Acknowledgements}

The authors gratefully acknowledge the anonymous referee for providing constructive comments and providing directions for additional work.
Based on service observations (program SW2014a08) made with the William Herschel Telescope (WHT), and on discretionary and Spanish CAT service observations made with the Isaac Newton Telescope (INT), both operated on the island of La Palma by the Isaac Newton Group (ING) in the Spanish Observatorio del Roque de los Muchachos of the Instituto de Astrof\'isica de Canarias. We thank the ING Director for having made public the INT data as soon as they were obtained. We also acknowledge the observers who kindly donated their time to monitor SN2014J on both the WHT and the INT.
We acknowledge Giuseppe Altavilla, Stefano Benetti for providing the data used in the left panel of Figure \ref{fig:alt} and Gast\'on Folatelli for the CSP data for Figure \ref{fig:alt}, \ref{fig:wang}, \ref{fig:gas2}, \ref{fig:gas1} and \ref{fig:ben}.
Support for LG and MH is provided by the Ministry of Economy, Development, and Tourism's Millennium Science Initiative through grant IC120009, awarded to The Millennium Institute of Astrophysics, MAS. LG acknowledges support by CONICYT through FONDECYT grant 3140566.
Support for MEMR and MM is provided by DGICYT grant AYA2010-21887-C04-02.
This work is partially funded by DGICYT grant AYA2013-47742-C4-4-P.
JIG-H, HL and JAR-M acknowledge financial support from the Spanish Ministry of Economy and Competitiveness (MINECO) under the 2011 Severo Ochoa program MINECO SEV-2011-0187
JIG-H also acknowledge the 2013 Ram\'on y Cajal program MINECO RYC-2013-14875, and the Spanish ministry project MINECO AYA2014-56359-P.
EAC acknowledges the support of the STFC.

\bibliographystyle{aa}
\bibliography{biblio}
\end{document}